\documentclass[10pt,a4paper]{article}
\usepackage{amsfonts}
\usepackage{amsthm,amsmath, amssymb}
\usepackage{graphicx}
\usepackage{verbatim}
\usepackage{booktabs}
\usepackage{caption}
\usepackage{natbib}
\usepackage{hyperref}
\usepackage{authblk}
\newcommand{\bo}[1]{\boldsymbol{#1}}
\newtheorem{theorem}{Theorem}
\newtheorem{definition}{Definition}
\newtheorem{remark}{Remark}
\newtheorem{corollary}{Corollary}
\newtheorem{lemma}{Lemma}

\newcommand{\bmat}{\begin{pmatrix}}
\newcommand{\emat}{\end{pmatrix}}
\usepackage{caption}
\usepackage{booktabs}

\newcommand\independent{\protect\mathpalette{\protect\independenT}{\perp}}
\def\independenT#1#2{\mathrel{\rlap{$#1#2$}\mkern2mu{#1#2}}}

\begin{document}

\title{Asymptotic and bootstrap tests for subspace dimension}


\author[1,2]{Klaus Nordhausen}
\author[1]{Hannu Oja \thanks{Corresponding author: E-mail hannu.oja@utu.fi}}
\author[3]{David E. Tyler}
\affil[1]{Department of Mathematics and Statistics, University of Turku}
\affil[2]{Institute of Statistics \& Mathematical Methods in Economics Vienna, University of Technology}
\affil[3]{Department of Statistics, The State University of New Jersey }







\maketitle

\begin{abstract}
Most linear dimension reduction methods proposed in the literature can be formulated using an appropriate pair of scatter matrices, see e.g. \citet{YeWeiss03}, \citet{TCDO09}, \citet{BuraYang11}, \citet{Liski2014} and \citet{LuoLi2016}. The eigen-decomposition of one scatter matrix with respect to another is then often used to determine the dimension  of the signal subspace and to separate signal and noise parts of the data. Three popular dimension reduction methods, namely principal component analysis (PCA), fourth order blind identification (FOBI) and sliced inverse regression (SIR) are considered in detail and the first two moments of subsets of the eigenvalues are used to test for the dimension of the signal space. The limiting null distributions of the test statistics are discussed and novel bootstrap strategies are suggested for the small sample cases. In all three cases, consistent test-based estimates of the signal subspace dimension are introduced as well.  The asymptotic and bootstrap tests are compared in simulations and illustrated in real data examples.
\end{abstract}



\section{Introduction}\label{sec:introduction}

Dimension reduction (DR) plays an increasingly  important role in high dimensional data analysis.
In linear dimension reduction for a random vector $\bo x\in \mathbb{R}^p$, the idea is to try to find a transformation matrix $\bo W\in  \mathbb{R}^{q\times p}$, $q\ll p$,  such that the interesting features of the distribution
of $\bo x$ are captured by $\bo W \bo x$ only,  that is,
\begin{itemize}
\item[(i)]\ \  $\bo x|\bo W\bo x$ is viewed as noise  (unsupervised DR), \ \ or
\item[(ii)]\ \   $ y \independent  \bo x\ |\ \bo W\bo x$ for the response of interest $ y$  (supervised DR).
\end{itemize}
In this paper we consider three classical but diverse linear dimension reduction methods:
principal component analysis, independent component analysis  and sliced inverse regression. As an introduction to our approach, we first
highlight the similarities between these three approaches. \\

Write $F_{\bo x}$ and $\bo S=\bo S(F_{\bo x})$ for the cumulative distribution function and covariance matrix of $\bo x$.
To simplify the notation, we assume in the following that $\mathbb{E}(\bo x)=\bo 0$.\\

(i) \ \
In the principal component analysis (PCA), one finds the $p\times p$ transformation matrix $\bo W$ such that
\[
\bo W \bo W'=\bo I_p
\ \ \ \ \mbox{and}\ \ \ \
\bo W \bo S \bo W'=\bo D
\]
where $\bo D$ is a diagonal matrix with diagonal elements $d_1\ge ...\ge d_p \ge 0$.
If $d_1\ge ...\ge d_q>d_{q+1}=...=d_p \ge 0$ and $\bo W$ is partitioned accordingly
as $\bo W=(\bo W_1',\bo W_2')'$, then $\bo W_1 \bo x$ is often seen as the $q$-variate signal part and $\bo W_2 \bo x$  as the $(p-q)$-variate noise part. Hence, $\bo W_2 \bo x$  is considered noise if and only if the eigenvalues of $\bo W_2 \bo S \bo W_2'$ are all equal.\\

 (ii) \ \
 In the independent component analysis (ICA) with $q$ non-Gaussian and $p-q$ Gaussian components, the fourth order blind identification (FOBI) method finds a transformation matrix $\bo W\in \mathbb{R}^{p\times p}$ such that
 \[
\bo W \bo S \bo W'=\bo I_p
\ \ \ \ \mbox{and}\ \ \ \
\bo W \mathbb{E}\left[\bo x \bo x' \bo S^{-1}\bo x\bo x'\right] \bo W'=\bo D
\]
where $\bo D$ is a diagonal matrix with the diagonal elements ordered so that $(d_1-(p+2))^2\ge ...  \ge (d_q-(p+2))^2$ $> (d_{q+1}-(p+2))^2=...=(d_p-(p+2))^2=0.$
Then $\bo W$  can again be partitioned as $\bo W=(\bo W_1',\bo W_2')'$ so that, under weak assumptions,  $\bo W_1 \bo x$ is the $q$-variate non-Gaussian signal  and $\bo W_2 \bo x$ the $(p-q)$-variate
Gaussian noise. If we further write
$\bo S_1:= \bo S$ {and} $\bo S_2:= \mathbb{E}\left[\bo x \bo x' \bo S_1^{-1}\bo x\bo x'\right]$
then,  $\bo W_2\bo x$ is considered noise if the eigenvalues of $\bo W_2 \bo S_2  \bo W_2'$ are all equal to  $p+2$.\\

(iii)\ \
In the sliced inverse regression (SIR)  with a $p$-variate random vector $\bo x$ and the response (dependent) variable $ y$,
one finds  a  matrix $\bo W\in \mathbb{R}^{p\times p} $ which satisfies
\[
\bo W \bo S_1  \bo W'=\bo I_p
\ \ \ \ \mbox{and}\ \ \ \
\bo W \bo S_2 \bo W'=\bo D
\]
where
$\bo S_1:= \bo S$  {and} $\bo S_2:= \mathbb{E}\left[ \mathbb{E}(\bo x | y) \mathbb{E}(\bo x |\bo y)'\right]$
and
 $\bo  D$ is a diagonal matrix with the diagonal elements  $d_1\ge ...\ge d_p\ge 0 $.
 Under appropriate assumptions on the distribution of $(\bo x,y)$, we have $d_1\ge ...\ge d_q> d_{q+1}=...=d_p=0 $ with the
 corresponding partitioning $\bo W=(\bo W_1',\bo W_2')'$. It is then thought that $(\bo W_1\bo x,y)$ carries all the information about the dependence between $\bo x$ and $y$,
 and $\bo W_2\bo x$ just presents noise.  Thus, $\bo W_2\bo x$ is thought to be noise if the eigenvalues of $\bo W_2 \bo S_2 \bo W_2'$ are all equal to zero.
 \\

To test and estimate the dimension of the signal space (also called order determination) and  to separate  signal and noise, we thus utilize, for appropriate choices of $\bo S_1$ and $\bo S_2$,  the  eigen-decomposition of $\bo S_1^{-1}\bo S_2$, or that of the symmetric matrix $\bo R:=\bo S_1^{-1/2}\bo S_2 \bo S_1^{-1/2}$.
 For the PCA case, we take $\bo S_1 = \bo I_p$  and $\bo S_2 = \bo S$, the covariance matrix, or some other scatter matrix, as defined later in Section~\ref{sec:scatter}.
 The tests are based on the first two moments of selected subsets of the eigenvalues of $\bo R$ 
and the corresponding estimates are
obtained applying different sequential testing strategies.  
The sequential testing procedures for the order determination  problem in SIR have been suggested earlier by \citet{Li91} and \citet{BuraCook01}.
\citet{Zhuetal06,Zhuetal10} used the  eigenvalues with BIC-type penalties to find consistent estimates
for the dimension of the signal subspace of a  regression model.
In other general approaches, \citet{YeWeiss03} considered eigenvectors rather than eigenvalues and proposed an estimation procedure that was based on the bootstrap variation of the subspace estimates for different dimensions.
In a  general approach, \citet{LuoLi2016}  combined the eigenvalues and bootstrap variation of eigenvectors for
consistent estimation of  the dimension. The last two approaches are based on the notion that the variation of eigenvectors is large for the the eigenvalues that are close  together
and their variability tends to be small for far apart eigenvalues.\\

In PCA the eigenvalues of $\bo S$  are generally used to make inference on the dimension of the signal space, see e.g. \citet{Jolliffe02} and  \citet{schott06} and references therein. Early papers on the use of bootstrap estimation and testing (via confidence intervals) in principal component analysis are \citet{BeranSrivastava85}, \citet{Daudin88}, \citet{EatonTyler91} and \citet{Jackson93}.
For the use of permutation tests in restricting the number of principal components, see
\citet{Dray08} and \citet{Vieira12}.\\

 In the independent component analysis (ICA) the fourth-order blind
identification (FOBI) by \citet{Cardoso89} uses the regular covariance matrix and the scatter matrix based on fourth moments and the eigenvalues provide measures of marginal kurtosis.
 These two matrices can be replaced by any two matrices possessing the so called independence property, see \citet{Oja:2006}, \citet{TCDO09} and \citet{Nordhausen:2015}.
 Very recently, \citet{Nordhausen:2017} used the the eigenvalues of $\bo S_1^{-1}\bo S_2$ to test and estimate the dimensions of Gaussian and non-Gaussian subspaces.\\

 PCA and FOBI are examples of {unsupervised dimension reduction} procedures as they do not use information on any response variable $y$. Other examples of unsupervised dimension reduction methods are invariant coordinate selection (ICS), see \citet{TCDO09}  , and generalized principal components analysis (GPCA), see \citet{CaussinusRuizGazen93}.
  Sliced inverse regression (SIR) uses the regular covariance matrix of $\bo x$ and the covariance matrix of the conditional expectation $E(\bo x|y)$. Other examples on {supervised dimension reduction} methods
are the canonical correlation analysis (CCA), sliced average variance estimate (SAVE) and principal Hessian directions (PHD), for example, and they all can be formulated using two scatter matrices. For these methods and estimation of the dimension of the signal subspace, also with regular bootstrap sampling,  see  \citet{Li91}, \citet{CookWeisberg91},
\citet{Li92}, \citet{BuraCook01}, \citet{Cook04}, \citet{Zhuetal06,Zhuetal10}, \citet{BuraYang11} and \citet{LuoLi2016}  and the references therein. For a nice review on supervised dimension reduction, see  \citet{MaZhu13}.
\\

The plan of this paper is as follows. In Section 2 we introduce the tools for our analysis, that is, the notion of a scatter matrix  with
some preliminary theory.
In all three cases in Sections 3 (PCA), 4 (FOBI) and 5 (SIR), respectively, we first specify a  natural semiparametric model: $\bo x=\bo A\bo z+\bo b$ where $\bo A$ and $\bo b$ are the parameters and the distribution of the standardized  $\bo z$ is  only partially specified. The null hypothesis says that $\bo z$ can be partitioned as $\bo z=(\bo z_1',\bo z_2')'$ and the first part $\bo z_1$  
carries the interesting variation. In the paper, the  eigenvalues of $\bo S_1^{-1}\bo S_2$, that is,
 the eigenvalues of $\bo R=\bo S_1^{-1/2}\bo S_2 \bo S_1^{-1/2}$, are utilized in this partitioning and  used to build tests and estimates for the dimension of $\bo z_1$. We discuss the asymptotic tests with corresponding estimates  and provide different strategies for bootstrap testing. Different approaches are compared in simulations and illustrated with real data examples. All the proofs are postponed to the Appendix.
\\

We adapt the following notation. $\mathbb{R}_{sym}^{p\times p}$ and $\mathbb{R}_{sym,+}^{p\times p}$ are the sets of symmetric and positive definite symmetric matrices, respectively.
The  first and second moments and the variance of the eigenvalues of $\bo R\in\mathbb{R}_{sym}^{p\times p}$ are denoted by
\[
m_1(\bo R) := tr(\bo R)/p, \ \  m_2(\bo R) :=  m_1(\bo R^2)\ \ \mbox{and}\ \ s^2(\bo R):= m_2(\bo R)-m_1^2(\bo R),
\]
respectively. If $\bo R=\bo U\bo D\bo U'\in\mathbb{R}_{sym,+}^{p\times p} $ is a eigen-decomposition of $\bo R$ then $\bo R^{1/2}:=\bo U\bo D^{1/2}\bo U'\in \mathbb{R}^{p\times q} $ (symmetric version of the square root matrix). Given $k$ matrices $\bo A_1,\bo A_2,...,\bo A_k$, we write
 \[
 diag(\bo A_1,...,\bo A_k)=\left(
                             \begin{array}{cccc}
                               \bo A_1 & \bo 0 & ... & \bo  0 \\
                               \bo 0 & \bo A_2 & ... & \bo 0 \\
                               ...& ...& ... & ...\\
                                \bo 0 & \bo 0 & ...  & \bo A_k\\
                             \end{array}
                           \right). 
 \]
 The vectorization of a matrix $\bo A\in \mathbb{R}^{p\times q}$, denoted by $vec(\bo A)$, is a $qp$-vector  obtained by stacking the columns of $\bo A$ on top of each other.  We further write  $\mathcal{O}^{p\times k}$, $k\le p$, for the set of column orthonormal  $p\times k$ matrices, i.e. $\bo U\in \mathcal{O}^{p\times k}$ implies $\bo U'\bo U=\bo I_k$.
 Hence, given $\bo U \in \mathcal{O}^{p \times k}$, $\bo P_{\bo U}:= \bo U\bo U'$ is the orthogonal projection onto the range of $\bo U$, and $\bo Q_{\bo U} = \bo I_p - \bo P_{\bo U}$
  is the orthogonal projection onto its orthogonal complement,  i.e.  onto the null space of $\bo U'$.
Let $\bo e_i\in \mathbb{R}^p$
denote the $i$th Euclidean basis element, i.e. a vector with a one in the $i$th position and zeroes elsewhere.
For two random vectors $\bo x$ and $\bo y$, we write $\bo x\sim \bo y$ if $\bo x$ and $\bo y$ has the same distribution.  The random vector $\bo z \in \mathbb{R}^p$ has a spherical distribution if $\bo U \bo z\sim \bo z$ for all $\bo U\in \mathcal{O}^{p\times p}$. The distribution of $\bo x$ is elliptical if $\bo x\sim \bo A \bo z +\bo b$ where $\bo A\in \mathbb{R}^{p\times p}$ and $\bo b\in \mathbb{R}^p$ and $\bo z\in \mathbb{R}$ has a spherical distribution.

\section{Scatter matrices }\label{sec:scatter}

In this chapter, we state what we ´mean by a scatter matrix and a supervised scatter matrix and provide some preliminary results.
Let $\bo F_{\bo x}$ be the cumulative distribution function (cdf) of a $p$-variate random vector $\bo x$ and  $\bo F_{\bo x, y}$ the cdf of the joint distribution of $p$-variate $\bo x$ and univariate $y$.

\begin{definition}\ \\
(i) The functional  $\bo S(F_{\bo x})\in\mathbb{R}^{p\times p}_{sym,+}$ is a scatter matrix (functional) if it is affine equivariant in the sense that
$\bo S(F_{\bo A\bo x+\bo b})=\bo A \bo S(F_{\bo x}) \bo A'$ for all non-singular $\bo A\in \mathbb{R}^{p\times p}$ and all $\bo b\in\mathbb{R}^p$.\\
(ii) The functional  $\bo S(F_{\bo x,y})\in\mathbb{R}^{p\times p}_{sym}$ is a supervised scatter matrix (functional) if it is affine equivariant in the sense that
$\bo S(F_{\bo A\bo x+\bo b,y})=\bo A \bo S(F_{\bo x, y}) \bo A'$ for all non-singular $\bo A\in \mathbb{R}^{p\times p}$ and all $\bo b\in\mathbb{R}^p$.
\end{definition}

Let $\bo X=(\bo x_1,...,\bo x_n)'\in \mathbb{R}^{n\times p}$ be a random sample from a distribution $F_{\bo x}$. The estimate  $\widehat{\bo S}$
of the population value  $\bo S(F_{\bo x})$ is obtained as the value of the functional at the empirical distribution $F_n$  of $\bo X$. We also write $\bo S(\bo X)$ for this estimate.
Let $\bo X=\bo Z\bo A'+\bo 1_n\bo b'$ where $\bo Z=(\bo z_1,...,\bo z_n)'$ is a random sample from a spherical distribution $F_{\bo z}$ with $\bo S(F_{\bo z})=\bo I_p$.
(Note that, for any scatter matrix $\bo S$,  $\bo S(F_{\bo z})\propto \bo I_p$ and can the rescaled to satisfy the last condition.)
Then $\bo X$ is a random sample from an elliptical distribution with $\bo S(F_{\bo x})=\bo A\bo A'$.\\

Under general assumptions,  the limiting distribution of
$\sqrt{n}\ vec({\bo  S}(\bo Z)-\bo I_p)$ is
\[
N_{p^2}\left(\bo 0, \sigma_1 (\bo I_{p^2}+\bo K_{p,p})+\sigma_2 vec(\bo I_p)vec(\bo I_p)'\right)
\]
where $\bo K_{p,p}=\sum_{i=1}^p\sum_{j=1}^p(\bo e_i\bo e_j')\otimes
(\bo e_j\bo e_i')$ is the commutation matrix, see Theorem 1 in \citet{tyler:1982}. The limiting distribution is known if the  following two constants, same for any $i\ne j$,
\[
\sigma_1=AsVar(\bo S (\bo Z)_{ij}),\ \ \mbox{and}\ \
\sigma_2=AsCov(\bo S (\bo Z) _{ii},\bo S (\bo Z) _{jj})
\] are known and then $AsVar(\bo S (\bo Z)_{ii})=2\sigma_1+\sigma_2$.
Also, under general conditions, the influence function of the scatter functional $\bo S(F)$ at a
spherical $F_{\bo z}$ is given by
\[
IF(\bo x;\bo  S,F_{\bo z})= \alpha(r) \bo u \bo u^T-\beta(r) \bo I_p
\]
where $r=||\bo x||$ and $\bo u=||\bo x||^{-1} \bo x$, see \citet{hampeletal:1986}.
If $\bo S(F)$ is the covariance matrix and $\bo S(F_{\bo z})=\bo I_p$,  then $\alpha(r)=r^2$ and $\beta(r)=1$ and if  $\bo z\sim N_p(\bo 0,\bo I_p)$ then  $\sigma_1=1$ and $\sigma_2=0$. For Tyler's shape estimate
(scaled so that its trace is $p$) which we use as a robust alternative in our simulations in Section~\ref{simulations:PCA}, $\alpha(r)=(p+2)$ and $\beta(r)=(p+2)/p$.\\

In the following we often need to estimate $\sigma_1$. It then follows, as noted in \citet{CH00}, that
$
\sigma_1={ E(\alpha^2(r))}/{(p(p+2))}
$.
Due to affine equivariance of the scatter matrix, the limiting distribution of  $\sqrt{n}\ vec({\bo  S}(\bo X)-\bo A\bo A')=\left(\bo A\otimes \bo A \right) \sqrt{n}\ vec({\bo  S}(\bo Z)-\bo I_p)$ and, using $\widehat{\bo S} $ with a companion location estimate $\hat {\bo\mu}$, $\sigma_1$ can often be consistently estimated by
\[
\hat\sigma_1= \frac 1 {p(p+2)} \frac 1n \sum_{i=1}^n \alpha^2(\hat
r_i)
\ \ \mbox{with}\ \
\hat r_i=\left((\bo x_i-\hat {\bo \mu})'\widehat {\bo S}^{-1}(\bo x_i-\hat
{\bo \mu})\right)^{1/2}.
\]
\\

\section{Testing for subspace dimension in PCA}\label{sec:PCA}

\subsection{The model, null hypothesis and test statistic}

Let $\bo X=(\bo x_1,...,\bo x_n)'$ be  a random sample from a $p$-variate elliptical distribution $F_{\bo x}$, that is, from the distribution of a random $p$-vector $\bo x$
generated by
\[
\bo x= \bo A \bo z+\bo b,
\]
where $\bo A\in \mathbb{R}^{p\times p}$ is non-singular, $\bo b\in \mathbb{R}^p$ and $\bo z$ has a spherical distribution around the origin, that is,
$\bo U\bo z\sim \bo z$ for all  $\bo U\in\mathcal{O} ^{p\times p}$. The distribution of $\bo z$ is then fully determined by the distribution
of its radius $r:=||\bo z||$. We assume that $\bo S(F_{\bo z})=\bo I_p$ for the scatter matrix functional  used in the analysis. For a general overview of spherical and elliptical distributions, see \citet{kelker70} or \citet{BB99}.
\\

If $\bo A=diag(\bo A_{11},a \bo I_{p-q})$   and the $q$ squared eigenvalues of $\bo A_{11}$ are larger than $a^2$,   we can write
\[
\bo x= \left(
         \begin{array}{c}
           \bo x_1 \\
           \bo x_2 \\
         \end{array}
       \right)
       =\left(
         \begin{array}{c}
           \bo A_{11}\bo z_1 \\
           a \bo z_2 \\
         \end{array}
       \right)+\bo b
\]
where $a\bo z_2\in \mathbb{R}^{p-q}$ is spherical.  we say that $\bo x$ is elliptical and subspherical around $\bo b$. Further, if
$\bo z$ is not spherical but $diag(I_q, \bo U) \bo z\sim \bo z$ for all $\bo U\in \mathcal{O}^{(p-q)\times (p-q)}$, then
 $diag(I_q, \bo U) (\bo x-\bo b)\sim (\bo x-\bo b)$ or all $\bo U\in \mathcal{O}^{(p-q)\times (p-q)}$ and $\bo x$ is said to be subspherical  around $\bo b$. The aim is to construct  tests and estimates for $q$ as well as to estimate the subvectors $\bo x_1\in \mathbb{R}^q$ (signal) and $\bo x_2\in \mathbb{R}^{p-q}$  (noise).\\

As the matrix of eigenvectors and the corresponding eigenvalues of  $\bo S(F_x)$ are equivariant and invariant, respectively,  under orthogonal transformations
 of $\bo x$, it is not a restriction to assume in our derivations that $\bo A$ is diagonal with positive and descending entries and $\bo b=\bo 0$ so that $\bo S(F_{\bo x})$ is a diagonal matrix $\bo D=\bo A^2$ with diagonal entries $d_1\ge ...\ge d_p > 0$. Let  $\widehat{\bo S}$ be the value of the scatter functional at the empirical distribution of $\bo X$.
 For the asymptotic results, we assume that  $\sqrt{n}vec(\widehat{\bo S}-\bo D)$ has a limiting multivariate normal distribution with zero mean vector and the covariance structure
 as described in Section~\ref{sec:scatter}. We wish to test the null hypothesis
\[
H_{0k}:\ \ d_1\ge ... \ge d_{k}>d_{k+1}=...=d_p=d\ \ \mbox{for
some unknown $d$,}
\]
stating that the dimension of the signal space is $k$. Under $H_{0k}$, the distribution of $\bo x$ is subspherical, that is, the distribution of the subvector of the last $p-k$
principal components is spherical. In principal component analysis, the scree plot is often used to figure out how many  components to include in the final  model. The null hypothesis $H_{0k}$ then implies that
there is the elbow on the scree plot at the $k$th eigenvalue. Also, sphericity and subsphericity (in a weaker sense) are important in the analysis of the repeated measures data, for example.  \\

To test the null hypothesis, we use  the variance of the $p-k$ smallest eigenvalues, that is,
\[
T_{k}:=  s^2(\widehat{\bo U}_k' \widehat{\bo S} \widehat{\bo U}_k) \ \ \mbox{with}\ \
\widehat{\bo U}_k=\arg \min_{\bo U\in \mathcal{O}^{p\times (p-k)}}  m_1(\bo U' \widehat{\bo S} \bo U)
\]
as a test statistic. It follows from the Poincar\'{e} separation theorem that a solution  $\widehat{\bo U}_k\in \mathcal{O}^{p\times(p-k)}$ is the matrix of
the eigenvectors associated with the $p-k$ smallest eigenvalues  of $\widehat{\bo S}$ and other solutions are obtained by post-multiplying it by an orthogonal  $(p-k)\times (p-k)$ matrix.
The projection matrices
$\widehat{\bo P}_k:=\widehat{\bo U}_k\widehat{\bo U}_k'$  and $\widehat{\bo Q}_k:=\bo I_p-\widehat{\bo P}_k$ are unique and
satisfy $\widehat{\bo P}_k\widehat{\bo S}\widehat{\bo Q}_k=\bo 0$
and provide the noise-signal decomposition $\bo x=\widehat{\bo P}_k \bo x+\widehat{\bo Q}_k \bo x$ with uncorrelated $\widehat{\bo P}_k \bo x$ and $\widehat{\bo Q}_k \bo x$.\\

Other possible measures for the variation of the smallest eigenvalues are the  coefficient of variation
$s(\widehat{\bo U}_k' \widehat{\bo S} \widehat{\bo U}_k)/m_1(\widehat{\bo U}_k' \widehat{\bo S} \widehat{\bo U}_k)$
or the log ratio of the arithmetic mean
$m_1(\widehat{\bo U}_k' \widehat{\bo S} \widehat{\bo U}_k)$ to the geometrical mean $det(\widehat{\bo U}_k' \widehat{\bo S} \widehat{\bo U}_k)^{1/(p-k)}$.
 If $\widehat{\bo S}$ is the covariance matrix, then the latter measure corresponds to
the likelihood ratio criterion for $H_{0k}$ in the multivariate normal case.

If one wishes to test a related null  hypothesis that $\bo S(F_{\bo x})$ has $k+1$ distinct eigenvalues with multiplicities $1,...,1,p-k$, then a natural
test statistic is 
\[
V_{k}:= \min_{\bo U\in \mathcal{O}^{p\times(p-k)}: \bo P_{\bo U}  \widehat{\bo S} \bo Q_{\bo U}=\bo 0 }  s^2\left(\bo U' \widehat{\bo S}\bo U\right).
\]
A solution $\widehat{\bo U}_k$ for which the minimum value is attained consists of the eigenvectors of $\widehat{\bo S}$ associated with the eigenvalues closest together
(in the variance sense). This is seen as follows. Let $\bo U\in \mathcal{O}^{p\times(p-k)}$ and  $\bo P_{\bo U}  \widehat{\bo S} \bo Q_{\bo U}=\bo 0$. Then  $\bo P_{\bo U}  \widehat{\bo S}= \widehat{\bo S} \bo P_{\bo U}$. As the symmetric matrices commute if and only if they have the same eigenvectors, $\bo U$ is a matrix of $p-k$ eigenvectors of $\widehat{\bo S}$, say $\bo U_0\in \mathcal{O}^{p\times(p-k)}$,
post-multiplied by an orthogonal $(p-k)\times (p-k)$ matrix.  Consequently, $\bo U' \widehat{\bo S}\bo U$ and $\bo U_0' \widehat{\bo S}\bo U_0$ have the same eigenvalues
and  $s^2(\bo U' \widehat{\bo S}\bo U)= s^2(\bo U_0' \widehat{\bo S}\bo U_0)$.  Thus the problem of minimizing $s^2(\bo U' \widehat{\bo S}\bo U)$ under the constraint
$\bo P_{\bo U}  \widehat{\bo S}\bo Q_{\bo U} = 0$ reduces to that of minimizing $s^2(\bo U_0' \widehat{\bo S}\bo U_0)$ over the $p-k$ subsets of eigenvectors of
$\widehat{\bo S}$.

\subsection{Asymptotic tests for dimension}

Assume now that $\bo x$ is elliptical with diagonal scatter matrix  $\bo D=\bo A^2$. Let $q$ denote the true value of the dimension of the signal space, that is, $H_{0q}$ is true, and consider the limiting distribution of
$T_q=s^2(\widehat{\bo U}_q' \widehat{\bo S}\widehat{\bo U}_q)$. With a correct value $q$ we have the partitions
\[
\bo D=\left(
              \begin{array}{cc}
                \bo D_1 & \bo 0 \\
                \bo 0 & d\bo I_{p-q}\\
              \end{array}
            \right)
\ \ \mbox{and}\ \
\widehat {\bo S}=\left(
   \begin{array}{cc}
    \widehat{\bo  S}_{11}  & \widehat{\bo  S}_{12} \\
     \widehat{\bo  S}_{21} & \widehat{\bo  S}_{22} \\
   \end{array}
 \right),
\]
respectively, and the diagonal elements in $\bo D_1$ are strictly larger than $d$. Under our assumptions, $\sqrt{n}(\widehat {\bo S}-\bo D)=O_P(1)$ and we have the following.

\begin{lemma}\label{L1:PCA}
Under the stated assumptions and $H_{0q}$,
$ n  T_q=n s^2(\widehat{\bo S}_{22}) +O_P(n^{-1/2}).
$
\end{lemma}

Under our assumptions stated  in Section~\ref{sec:scatter},  $\sqrt{n}\ vec({\bo  S}(\bo Z)-\bo I_p)$ where $\bo Z=\bo X\bo D^{-1/2}$ converges in distribution to a $p^2$-variate normal distribution
with zero mean vector and the covariance matrix $\sigma_1 (\bo I_{p^2}+\bo K_{p,p})+\sigma_2 vec(\bo I_p)vec(\bo I_p)'$.
Then we have the following.

\begin{theorem}\label{Th1:PCA}
Under the previously stated assumptions and under  $H_{0q}$, 
 $$\frac {n(p-q) T_q } {2 d^2\sigma_1} \to_d\chi^2_{\frac 12 (p-q-1)(p-q+2)}.$$
If the multiplicities of the eigenvalues of $\bo D_1$ are smaller than $p-q$ then $P(V_q=T_q)\to 1$ and the limiting distributions of $nV_q$ and $nT_q$ are the same.
\end{theorem}

 For the test construction in practice we thus need to estimate two  population constants $\sigma_1$ and $d$, both of which are invariant under orthogonal transformations to $\bo x$.
 The limiting distribution in Theorem~\ref{Th1:PCA} stays the same even if $\sigma_1$ and $d$ are replaced by their consistent estimates, say $\hat\sigma_1$ and $\hat d$. Construction of a consistent estimate for $\sigma_1$  has already been discussed  in Section~\ref{sec:scatter}. The unknown $d$ can be consistently estimated by the average of the $p-q$ smallest eigenvalues, that is, by $\hat d=m_1(\widehat{\bo U}_q' \widehat{\bo S}\widehat{\bo U}_q)$.   Note also that the test statistic in Theorem \ref{Th1:PCA} with these replacements depends on the smallest eigenvalues through their coefficient of variation, a test statistic suggested by \citet{schott06}. As noted previously, a possible test statistic for $H_{0q}$ is also the log of the ratio of the arithmetic and geometric means of the smallest $p-q$ eigenvalues of $\widehat{\bo S}$, say $L_q$.
Then under the null hypotheses as well as  under certain contiguous alternatives, $n(T_q - 2d^2 L_q) \to_p 0$
and then, under $H_{0q}$,  $n(p-q) L_q/\hat{\sigma}_1 \to_d\chi^2_{(p-q-1)(p-q+2)/2}$. See Theorem 5.1 and 5.2 and their proofs in \citet{tyler:1983}.

We now utilize the test statistics $T_k$, $k=0,1,...,p-1$, for the estimation problem
and collect some useful  limiting properties in the following theorem.

\begin{theorem}\label{Th2:PCA}
Under the previously stated assumptions and under $H_{0q}$,
 \begin{itemize}
\item[(i)] for $k<q$, $T_k\to_P c_k$ for some $c_1,...,c_{q-1}>0$,
\item[(ii)] for $k=q$,
${n(p-q) T_q }/{(2 d^2\sigma_1)} \to_d\chi^2_{\frac 12 (p-q-1)(p-q+2)}  $,
and
\item[(iii)] for $k>q$,
$nT_k \le  (\frac{p-q}{p-k})^2 nT_q=O_P(1)$.
\end{itemize}
\end{theorem}



A consistent estimate $\hat q$ of the unknown dimension  $q\le p-1$ can  then  be based on the test statistics $T_k$, $k=0,...,p-1$, as follows.
\begin{corollary}
For all $k=0,...,p-1$,  let
  $(c_{k,n})$  be a sequence of positive real numbers such that
$c_{k,n}\to 0$ and $n {c_{k,n}}\to \infty$ as $n\to\infty$.
Then, under the assumptions of Theorem~\ref{Th2:PCA},
\[
\mathbb{P}(T_k\ge c_{k,n})\ \rightarrow\
\left\{
       \begin{array}{ll}
         1, & \hbox{if $k<q$ ;} \\
         0, & \hbox{if $k\ge q$.}
       \end{array}
     \right.
\]
and
$\hat q=\min \{k\ :\ T_k < c_{k,n}\}\to_P q$. 
\end{corollary}

Note that, by definition,  $T_{p-1}=0$ and the maximum value of $q$ is  $p-1$, which corresponds to the smallest eigenvalue being distinct. The estimate  $\hat q$ is easily found by using the so called bottom-up testing strategy:  Start with tests for $H_{00}$, $H_{01}$ and so on, and stop when you get the first acceptance. An alternative consistent estimate with a top-down testing strategy is  $\hat q=\max \{k\ :\ T_{k-1} \ge c_{k-1,n}\}$  using successive tests for $H_{0,p-2}, H_{0,p-3},...,$ and stopping after the first rejection. For large $p$, faster strategies such as the divide and conquer algorithm are naturally available in the estimation.

 Let $F_k$ be the limiting distribution of $nT_k$ under $H_{0k}$.
 The sequences of critical values $(c_{k,n})$  for testing $H_{0k}$
can be determined by the corresponding  sequences of asymptotical  test sizes $(\alpha_{k,n})$
satisfying $\alpha_{k,n}=1-F_k(nc_{k,n})$
A simple and practical choice of the sequences of  the test sizes is for example $\alpha_{k,n} = (n_0/n)\alpha_k$, $k\le p-2$ and  $n\ge n_0$.
Then  $nc_{k,n}\to\infty$ as $\alpha_{k,n}=1-F_k(nc_{k,n})\to 0$,  and  $c_{k,n}\to 0$ as  $nc_{k,n} \alpha_{k,n}=  nc_{k,n} (1-F_k(n c_{k,n})) \to 0$.

To end the discussion on asymptotics, suppose we relax now the ellipticity assumption and consider a  model for which  
$diag(\bo I_q, \bo U) \bo z\sim \bo z$ for all $\bo U\in \mathcal{O}^{(p-q)\times (p-q)}$.
Since $\bo D=\bo A^2=diag(\bo D_1,d\bo I_{p-q})$,  $\bo x$ is subspherical but not necessarily elliptical. It is then easy to show that, for the covariance matrix
and finite fourth moments,
Lemma~\ref{L1:PCA} and Theorem~\ref{Th1:PCA} still hold true with  $\sigma_1=1$.  For other scatter matrices, however,  the asymptotic behavior in this wider model
is not known. \\

 Lemma~\ref{L1:PCA}  shows the remarkable  fact that under the null hypothesis $H_{0q}$ the limiting distributions of $nT_q=ns^2(\widehat{\bo U}_q' \widehat{\bo S}\widehat{\bo U}_q)$ and that of $ns^2({\bo U}_q' \widehat{\bo S}{\bo U}_q)$  with known noise subspace are the same. If, in the small sample case, the  $p$-values are obtained from the limiting  distribution of
 the test statistic, the variation coming from the estimation of the subspace is thus ignored in the null asymptotic approximation. In the following we therefore propose that the small sample null distribution of a test statistic be estimated by resampling the data from a distribution obeying the null hypothesis and being as close as possible to the empirical distribution.

\subsection{Bootstrap tests for dimension}

Again, let  $q$ denote the true dimension of the signal space and we wish to test the null hypothesis
\[
H_{0k}:\ \ d_1 \geq ... \geq d_{k}>d_{k+1}=...=d_p=d\ \ \mbox{for
some $d$.}
\]
It is important to stress that, in the practical testing situation,  we do not know whether  $H_{0k}$ is true ($k=q$) or whether it is false ($k\ne q$) but we still wish to
compute the $p$-values for true $H_{0k}$. See \citet{HallWilson91} for some guidelines in bootstrap hypothesis testing.
For testing, we start with a scatter matrix estimate $ \widehat {\bo S}$ and a companion location estimate $\widehat{\bo\mu}$ and compute
$
\widehat{\bo U}_k
$
and
$ T_k=s^2(\widehat{\bo U}_k'\widehat{\bo S}\widehat{\bo U}_k)$, the variance of $p-k$ smallest eigenvalues of $\widehat{\bo S}$. We further write
$\widehat{\bo P}_k= \widehat{\bo U}_k\widehat{\bo U}_k'$ and
$\widehat{\bo Q}_k=\bo I_p-\widehat{\bo P}_k$
for the estimated projection matrices to the noise and signal subspace under true $H_{0k}$, respectively.

The basic idea in the bootstrap testing strategy is that the bootstrap samples $\bo X^*$ for $H_{0k}$ should be generated from a distribution $F_{n,k}$
\begin{itemize}
\item[(i)] for which  the null hypothesis $H_{0k}$ is true (even if $k\ne q$) and
\item[(ii)] which is as close as possible
to the empirical distribution $F_n$ of $\bo X$.
\end{itemize}
 We suggest  the following two procedures. In the first procedure, the bootstrap samples come from a subspherical and  elliptical distribution (with the distribution of the radius estimated from the data)  and, in the second procedure,
 they come  a subspherical distribution (not assuming full ellipticity).  It is important that the dimension of the subspherical
 part is $p-k$ even when $k\ne q$.
 If one wishes to  assume multivariate normality then the first procedure can be further modified accordingly.  \\

{\bf Bootstrap strategy PCA-I (elliptical subspherical distribution):}\ \
 {\it
\begin{enumerate}
\item Starting with $\bo X\in \mathbb{R}^{n\times p}$, compute $\widehat{\bo \mu} $, $\widehat{\bo S}$ with the estimated matrix of  eigenvectors in $\widehat{\bo U}$ and
corresponding estimated eigenvalues in $\widehat{\bo D}$.
 \item Take a bootstrap sample   $\widetilde{\bo Z}=(\tilde{\bo z}_1, \ldots, \tilde{\bo z}_n)'$ of size $n$  from
 $(\bo X-\bo 1_n\widehat{\bo \mu}')\widehat{\bo U}\widehat{\bo D}^{-1/2} $.
\item For ellipticity  to be true, transform
\[ \bo z_i^*=\bo O_i \tilde{\bo z}_i,\ \ i=1,...,n,  \]
and  $\bo O_1, \ldots, \bo O_n\in  \mathcal{O}^{p\times p} $ are iid from the   Haar distribution.
 \item For subsphericity to be true as well, the bootstrap sample is $$\bo X^* = \bo Z^* \widehat{\bo D}_k ^{1/2} \widehat{\bo U}'+ \bo 1_n\widehat{\bo \mu}'$$
  where $\widehat{\bo D}_k=diag(\hat d_1,...,\hat d_k, \sum_{i=k+1}^p \hat d_i/(p-k),...,\sum_{i=k+1}^p \hat d_i/(p-k))$.
\end{enumerate}}

{\bf  Bootstrap strategy PCA-II (subspherical distribution):}\ \
{\it
\begin{enumerate}
\item Starting with  $\bo X\in \mathbb{R}^{n\times p}$, compute $\widehat{\bo S}$, $\widehat{\bo\mu}$, $\widehat{\bo U}_k$, $\widehat{\bo P}_k$  and $\widehat{\bo Q}_k$.
 \item Take a bootstrap sample   $\widetilde{\bo X}=(\tilde{\bo x}_1, \ldots, \tilde{\bo x}_n)'$ of size $n$  from  $\bo X$.
\item For subsphericity to be true, transform
\[ \bo x_i^*=\left[ \widehat{\bo Q}_k+\widehat{\bo U}_k \bo O_i \widehat{\bo U}_k'   \right] (\tilde{\bo x}_i-\widehat{\bo \mu}) +\widehat{\bo \mu},\ \ i=1,...,n,  \]
and  $\bo O_1, \ldots, \bo O_n\in  \mathcal{O}^{(p-k)\times (p-k)} $ are iid from the  Haar distribution.
 \item The bootstrap sample is $\bo X^* = (\bo x_1^*,...,\bo x_n^*) $.
\end{enumerate}}

 For both strategies and for  $k=0,...,p-1$,  the hypothesis $H_{0k}$ is true for the corresponding bootstrap null distribution, say $F_{n,k}$.
For the PCA-I strategy,
\begin{eqnarray*}
F_{n,k}(\bo x) &=&\frac 1n \sum_{i=1}^n \mathbb{E}_{\bo O_{i,p}} \left[ \mathbb{I}\left(\widehat{\bo U}_k \widehat{\bo D}_k^{1/2} \bo O_{i,p}\widehat{\bo D}^{-1/2}\widehat{\bo U}_k'
({\bo x}_i-\widehat {\bo \mu}) +\widehat {\bo \mu}
 \le \bo x  \right) \right]
\end{eqnarray*}
with random matrices $\bo O_{1,p},...,\bo O_{n,p}\in \mathcal{O}^{p\times p}$ from the Haar distribution.
Similarly, for the PCA-II strategy,
\begin{eqnarray*}
F_{n,k}(\bo x) &=&\frac 1n \sum_{i=1}^n \mathbb{E}_{\bo O_{i,p-k}} \left[ \mathbb{I}\left( (\widehat{\bo Q}_k+\widehat{\bo U}_k \bo O_{i,p-k} \widehat{\bo U}_k') ({\bo x}_i-\widehat {\bo \mu}) +\widehat {\bo \mu}
 \le \bo x  \right) \right]
\end{eqnarray*}
where $O_{1,p-k},...,O_{n,p-k}\in \mathcal{O}^{(p-k)\times (p-k)}$ are from the Haar distribution.  \\

Consider next the distribution of $nT_k(\bo X^*)$ for the PCA-I strategy.
Let then $\bo X^*_N\in \mathbb{R}^{N\times p}$ be a random sample of size $N$ from  $F_{n,k}$.
Note that $F_{n,k}$  is an elliptical distribution with true $H_{0k}$ and with data dependent parameters, namely, symmetry center $\bo \mu:=\widehat{\bo\mu}$, covariance matrix $\bo S:= \widehat{\bo U}\widehat{\bo D}_k \widehat{\bo U}'$ and
\[
d:=\hat d =\frac 1{p-k} \sum_{i=k+1}^p \hat d_i \ \ \mbox{and}\ \
\sigma_1:= \hat\sigma_1=\frac 1 {p(p+2)} \frac 1n \sum_{i=1}^n \alpha^2(\hat
r_i)
\]
where
$\hat r_i=((\bo x_i-\hat {\bo \mu})'\widehat {\bo S}^{-1}(\bo x_i-\hat
{\bo \mu}))^{1/2}$, $i=1,...,n$. Theorem~\ref{Th1:PCA} then implies that, given $\bo X$,  $N(p-k) T_k(\bo X_N^*) /(2 \hat d^2\hat \sigma_1)\to_d\chi^2_{\frac 12 (p-k-1)(p-k+2)}$ (a.s.) which provides,
for large $n$, the same asymptotic  chi-squared approximation for the distribution of the unconditional $n (p-k) T_k(\bo X^*) /(2 \hat d^2\hat \sigma_1)$ as well. Theorem~\ref{Th1:PCA}
 gave the same approximation for
$n (p-k) T_k(\bo X) /(2 \hat d^2\hat \sigma_1)$.
For the PCA-I strategy applied to the covariance matrix, similar arguments can be used to get  the same approximations for the distributions of $n (p-k) T_k(\bo X^*) /(2 \hat d^2)$
and  $n (p-k) T_k(\bo X) /(2 \hat d^2)$.
\\

In practice, the exact $p$-values are not computed  but estimated as follows.
Let $T=T(\bo X)$ be a test statistic for $H_{0k}$ such as  $T_k$, that is, the variance of the $p-k$ smallest eigenvalues of $\widehat{\bo S}$. If $\bo X_{1}^*,...,\bo X_{M}^*$ are independent bootstrap samples of size $n$ as described above and
$T_i^*=T(\bo X_{i}^*)$, $i=1,...,M$,  then the bootstrap $p$-value is given by
\[
\hat p= \frac {\# (T_i^*\ge T)+1} {M+1}.
\]
Note that, conditioned on $\bo X$,  $\hat p$ is a random variable whose variance around the true $p$-value can be estimated by $\frac 1M \hat p(1-\hat p)$.

\subsection{A simulation study}\label{simulations:PCA}

In the simulation study for the bootstrap tests
we wish to estimate the unknown rejection probability with a nominal level $\alpha$ at any distribution $F$.
For the estimation we use $N$ repetitions, that is, $N$ independent random samples
$\bo X_1,...,\bo X_N\in \mathbb{R}^{n \times p}$ from $F$ and, for each repetition,  we generate  $M$ bootstrap samples
denoted by $\bo X_{i1}^*,...,\bo X_{iM}^* \in \mathbb{R}^{n \times p} $, $i=1,...,N$. The observed bootstrap $p$-values then are
\[
 \hat p_{i}=\frac  { \sum_{j=1}^M \mathbb{I}(T(\bo X_{ij}^*)\ge T(\bo X_i))+1}{M+1}  ,\ \ i=1,...,N.
\]
For the $i$th sample $\bo X_i$, the null hypothesis is rejected  if $\hat p_{i}\le \alpha$ and the estimated rejection rate based on $\bo X_1,...,\bo X_N$
is
\[
\hat \beta =\frac 1 N  \sum_{i=1}^N \mathbb{I}\left(\hat p_{i}\le \alpha  \right).
\]
Then $\hat\beta$ is unbiased estimate of $\beta=\mathbb{P}\left(\hat p_{i} \le \alpha  \right)$,  the power at $F$,
which slightly depends on $M$ and its variance has an upper limit $\frac 1{4N}$.
 For our choice $N=2000$, the upper limit for the standard deviation is then  0.011.
 In our simulations we use $M=200$ but a larger value of $M$, say $M=500$, is naturally recommended in an analysis of a single data set.\\


The problem of dimension reduction with PCA often arises in the case of a latent factors model $\bo x = \bo A\bo z + \bo \epsilon$, where $\bo A\in \mathbb{R}^{p \times q}$, $q<p$,
is a matrix of loadings,  the latent random vector  $\bo z\in \mathbb{R}^q$ is seen as the signal, and $\bo \epsilon$ independent of $\bo z$  is an additive $p$-variate noise. An  example is a classical factor analysis model with $\bo z \sim N(\bo 0, \bo I_q)$ and $\bo \epsilon \sim N(\bo 0, \sigma^2 \bo I_p)$ (equal uniqueness for all marginal variables).
The so called noisy independent component (ICA) model is obtained if $\bo z$ has mutually independent non-Gaussian components and again $\bo \epsilon \sim N(\bo 0, \sigma^2 \bo I_p)$.
In both cases the noise part is subspherical and we wish to make inference on the dimension $q$ of the signal space.
In our simulations, we have the following three simulation settings with $q=3$ and  dimensions $p=6$ and $p=15$.
\begin{description}
  \item[M1:]  The factor analysis model with $\bo A\in \mathbb{R}^{p \times 3}$ having the only three non-zero elements $a_{11}=\sqrt{2}$ and $a_{22}=a_{33}=1$.
  $\sigma^2=1$.
  \item[M2:] A noisy ICA model with $\bo A\in \mathbb{R}^{p \times 3}$ having the only three non-zero elements $a_{11}=\sqrt{2}$ and $a_{22}=a_{33}=1$.
   $\bo z$ has the three standardized independent components,  exponential, $\chi^2_1$ and $t_5$. $\sigma^2=1$.
  \item[M3:] An elliptical $p$-variate $t_5$ distribution with $Cov(\bo x) = diag(3,2,2,1,\ldots,1)$.
\end{description}

In all three models the covariance matrix of $\bo x$ is $diag(3,2,2,1,\ldots,1)$   and  models M1 and M3 state an elliptical distribution. The sample covariance
matrix is optimal only for the model M1. 
We also use  the Tyler's shape matrix \citep{tyler:1987} together with the companion location estimate, the spatial median; the pair of the estimates is called the Hettmansperger-Randles (HR) estimate \citep{hettmanspergerandrandles:2002}. The HR estimate is expected to be more efficient for heavy-tailed distributions such as the $t_5$ distribution in
 the model M3. Tyler's shape matrix is diagonal in models M1 and M3 but only block-diagonal  in M2 but the $p-q$  smallest eigenvalues are equal in all cases, see \citet{Nordhausen:2015}. For a  discussion of robustness and computational issues in bootstrapping
see for example \citet{saliban:2002,saliban:2005}.
In general,  the computation of the M-estimators such as Tyler's shape matrix is fast  for  a single data set,
see e.g. \citet{Duembgen:2016}. In our simulations  we however adopt, in the spirit of fast and robust bootstrap,
a 3-step  fixed-point estimates \citep{Taskinen:2016} for the bootstrap samples $\bo X_{ij}^*$ starting with an estimate from the original data set $\bo X_i$ and
utilizing the bootstrap sample $\bo X_{ij}^*$ when updating the estimate three times. \\

In the simulation we use the asymptotic and bootstrap (strategies PCA-I and PCA-II) tests that use the covariance matrix and Tyler's shape matrix.
Schott's asymptotic test \citep{schott06} (with a finite-sample correction) is used as a standard reference test although it is expected to work well only under the model M1.
The estimated rejection rates for $\alpha=0.05$  for false $H_{02}$ and true $H_{03}$ based on $N=2000$ repetitions are reported for the dimensions $p=6,15$ and
sample sizes  $n= 50,100,500,1000$.   (In all cases, the rejection rates for $H_{04}$ tend to be much smaller than those for $H_{03}$ and are not reported here.)
In M1 we computed both the Tyler's original shape matrix and its 3-step version in the bootstrapping. \\

\begin{table}[ht]
\centering
\tiny
\begin{tabular}{@{\extracolsep{4pt}}llccccccccc}
  \hline
p  & n  & Schott & \multicolumn{3}{c}{Covariance matrix} & \multicolumn{5}{c}{Tyler's shape matrix}\\
   &    &        & Asymp & PCA-I & PCA-II & Asymp & PCA-I & PCA-I & PCA-II & PCA-II \\
   &    &        &  &  &  &  & full & 3-step & full & 3-step \\ \cline{3-3} \cline{4-6} \cline{7-11}
  6 & 50    & 0.2295 & 0.3530 &0.2965 &0.3010 &  0.1790& 0.1810 & 0.2365 & 0.1695 & 0.2205 \\
   & 100    & 0.6560 & 0.7470 &0.7085 &0.7190 &  0.5295& 0.5370 & 0.5785 & 0.5310 & 0.5685 \\
   & 500    & 1.0000 & 1.0000 &1.0000 &1.0000 &  1.0000& 1.0000 & 1.0000 & 1.0000 & 1.0000 \\
   & 1000   & 1.0000 & 1.0000 &1.0000 &1.0000 &  1.0000& 1.0000 & 1.0000 & 1.0000 & 1.0000 \\   \hline
 15 & 50    & 0.0687 & 0.2867 &0.1067 &0.1230 &  0.2290& 0.0920 & 0.1903 & 0.0910 & 0.1873 \\
  & 100     & 0.2510 & 0.5130 &0.3850 &0.4010 &  0.3960& 0.3130 & 0.3650 & 0.3120 & 0.3670 \\
  & 500     & 1.0000 & 1.0000 &1.0000 &1.0000 &  1.0000& 1.0000 & 1.0000 & 1.0000 & 1.0000 \\
  & 1000  & 1.0000 & 1.0000 &1.0000 &1.0000 &  1.0000& 1.0000 & 1.0000 & 1.0000 & 1.0000 \\
   \hline
\end{tabular}
\caption{Rejection rates for false $H_{02}$  in the multivariate normal case (M1).}
\label{PCA_M1_k2}
\end{table}

\begin{table}[ht]
\centering
\tiny
\begin{tabular}{@{\extracolsep{4pt}}llccccccccc}
  \hline
p  & n  & Schott & \multicolumn{3}{c}{Covariance matrix} & \multicolumn{5}{c}{Tyler's shape matrix}\\
   &    &        & Asymp & PCA-I & PCA-II & Asymp & PCA-I & PCA-I & PCA-II & PCA-II \\
   &    &        &  &  &  &  & full & 3-step & full & 3-step \\ \cline{3-3} \cline{4-6} \cline{7-11}
  6 & 50   & 0.0375 & 0.0590& 0.0460 & 0.0500 & 0.0400 & 0.0500 & 0.0660& 0.0445 & 0.0635 \\
   & 100   & 0.0500 & 0.0635& 0.0530 & 0.0570 & 0.0480 & 0.0540 & 0.0760& 0.0520 & 0.0695 \\
   & 500   & 0.0470 & 0.0505& 0.0395 & 0.0495 & 0.0565 & 0.0560 & 0.0670& 0.0545 & 0.0645 \\
   & 1000  & 0.0485 & 0.0510& 0.0425 & 0.0495 & 0.0455 & 0.0465 & 0.0530& 0.0445 & 0.0485 \\   \hline
 15 & 50   & 0.0173 & 0.0843& 0.0143 & 0.0233 & 0.0667 & 0.0150 & 0.0480& 0.0180 & 0.0463 \\
  & 100    & 0.0260 & 0.0710& 0.0360 & 0.0420 & 0.0710 & 0.0430 & 0.0650& 0.0460 & 0.0670 \\
  & 500    & 0.0470 & 0.0550& 0.0435 & 0.0510 & 0.0525 & 0.0490 & 0.0540& 0.0490 & 0.0540 \\
  & 1000 & 0.0470 & 0.0510& 0.0415 & 0.0485 & 0.0450 & 0.0455 & 0.0525& 0.0420 & 0.0510 \\
   \hline
\end{tabular}
\caption{Rejection rates for true $H_{03}$  in the multivariate normal case (M1).}
\label{PCA_M1_k3}
\end{table}

\begin{table}[ht]
\centering
\tiny
\begin{tabular}{@{\extracolsep{4pt}}llccccccc}
  \hline
p  & n  & Schott & \multicolumn{3}{c}{Covariance matrix} & \multicolumn{3}{c}{Tyler's shape matrix}\\
   &    &        & Asymp & PCA-I & PCA-II & Asymp & PCA-I  & PCA-II \\
  \cline{3-3} \cline{4-6} \cline{7-9}
  6 & 50  & 0.2040 & 0.2510 & 0.2340 & 0.2385& 0.1040 & 0.1505 & 0.1445\\
   & 100  & 0.5740 & 0.5895 & 0.6030 & 0.6150& 0.3055 & 0.3620 & 0.3595\\
   & 500  & 1.0000 & 1.0000 & 1.0000 & 1.0000& 0.9965 & 0.9955 & 0.9965\\
   & 1000 & 1.0000 & 1.0000 & 1.0000 & 1.0000& 1.0000 & 1.0000 & 1.0000\\    \hline
 15 & 50  & 0.0590 & 0.2505 & 0.1020 & 0.1215& 0.1750 & 0.1450 & 0.1355\\
  & 100   & 0.2195 & 0.4370 & 0.3495 & 0.3545& 0.2825 & 0.2680 & 0.2610\\
  & 500   & 0.9985 & 1.0000 & 1.0000 & 0.9995& 0.9930 & 0.9925 & 0.9930\\
  & 1000  & 1.0000 & 1.0000 & 1.0000 & 1.0000& 1.0000 & 1.0000 & 1.0000\\
   \hline
\end{tabular}
\caption{Rejection rates for false $H_{02}$ in the  noisy ICA case (M2).}
\label{PCA_M2_k2}
\end{table}

\begin{table}[ht]
\centering
\tiny
\begin{tabular}{@{\extracolsep{4pt}}llccccccc}
  \hline
p  & n  & Schott & \multicolumn{3}{c}{Covariance matrix} & \multicolumn{3}{c}{Tyler's shape matrix}\\
   &    &        & Asymp & PCA-I & PCA-II & Asymp & PCA-I &  PCA-II \\
 \cline{3-3} \cline{4-6} \cline{7-9}
  6 & 50  & 0.0370 & 0.0415& 0.0340 & 0.0460 & 0.0235 & 0.0530& 0.0455 \\
   & 100  & 0.0420 & 0.0335& 0.0360 & 0.0455 & 0.0340 & 0.0575& 0.0530 \\
   & 500  & 0.0505 & 0.0320& 0.0380 & 0.0550 & 0.0495 & 0.0610& 0.0555 \\
   & 1000 & 0.0535 & 0.0320& 0.0370 & 0.0535 & 0.0545 & 0.0620& 0.0620 \\          \hline
 15 & 50  & 0.0145 & 0.0785& 0.0160 & 0.0235 & 0.0510 & 0.0380& 0.0390 \\
  & 100   & 0.0235 & 0.0530& 0.0310 & 0.0375 & 0.0385 & 0.0470& 0.0405 \\
  & 500   & 0.0470 & 0.0435& 0.0475 & 0.0505 & 0.0645 & 0.0685& 0.0605 \\
  & 1000  & 0.0470 & 0.0335& 0.0405 & 0.0430 & 0.0580 & 0.0575& 0.0510 \\
   \hline
\end{tabular}
\caption{Rejection rates for true $H_{03}$  in the noisy ICA case (M2).}
\label{PCA_M2_k3}
\end{table}

\begin{table}[ht]
\centering
\tiny
\begin{tabular}{@{\extracolsep{4pt}}llccccccc}
  \hline
p  & n  & Schott & \multicolumn{3}{c}{Covariance matrix} & \multicolumn{3}{c}{Tyler's shape matrix}\\
   &    &        & Asymp & PCA-I & PCA-II & Asymp & PCA-I  & PCA-II \\
  \cline{3-3} \cline{4-6} \cline{7-9}
  6 & 50  & 0.4170 & 0.2525& 0.3015& 0.2865  & 0.1835 & 0.2445& 0.2320 \\
   & 100  & 0.7730 & 0.4570& 0.5575& 0.5460  & 0.5380 & 0.6050& 0.5920\\
   & 500  & 1.0000 & 0.9910& 0.9965& 0.9985  & 1.0000 & 1.0000& 1.0000\\
   & 1000 & 1.0000 & 1.0000& 1.0000& 1.0000  & 1.0000 & 1.0000& 1.0000\\    \hline
 15 & 50  & 0.6890 & 0.6605& 0.6420& 0.2895  & 0.2025 & 0.1645& 0.1595\\
  & 100   & 0.9595 & 0.6945& 0.8360& 0.4890  & 0.4155 & 0.3915& 0.3920\\
  & 500   & 1.0000 & 0.9955& 0.9985& 0.9790  & 1.0000 & 1.0000& 1.0000\\
  & 1000  & 1.0000 & 1.0000& 1.0000& 0.9990  & 1.0000 & 1.0000& 1.0000\\
   \hline
\end{tabular}
\caption{Rejection rates for false $H_{02}$ in the  $t_5$ elliptical case (M3).}
\label{PCA_M3_k2}
\end{table}

\begin{table}[ht]
\centering
\tiny
\begin{tabular}{@{\extracolsep{4pt}}llccccccc}
  \hline
p  & n  & Schott & \multicolumn{3}{c}{Covariance matrix} & \multicolumn{3}{c}{Tyler's shape matrix}\\
   &    &        & Asymp & PCA-I & PCA-II & Asymp & PCA-I &  PCA-II \\
 \cline{3-3} \cline{4-6} \cline{7-9}
  6 & 50  & 0.1290 & 0.0575& 0.0855& 0.0895  & 0.0320& 0.0580& 0.0560  \\
   & 100  & 0.2060 & 0.0530& 0.0945& 0.0950  & 0.0380& 0.0605& 0.0605  \\
   & 500  & 0.3520 & 0.0550& 0.0755& 0.0795  & 0.0500& 0.0565& 0.0570  \\
   & 1000 & 0.4195 & 0.0640& 0.0750& 0.0780  & 0.0500& 0.0630& 0.0600  \\          \hline
 15 & 50  & 0.4190 & 0.3770& 0.3500& 0.1665  & 0.0650& 0.0455& 0.0465  \\
  & 100   & 0.7515 & 0.2655& 0.4895& 0.2270  & 0.0490& 0.0485& 0.0445  \\
  & 500   & 0.9765 & 0.1910& 0.4030& 0.1855  & 0.0525& 0.0545& 0.0525  \\
  & 1000 & 00.9905 & 0.1630& 0.2920& 0.1365  & 0.0470& 0.0485& 0.0480 \\
   \hline
\end{tabular}
\caption{Rejection rates for true $H_{03}$   $t_5$ elliptical case (M3).}
\label{PCA_M3_k3}
\end{table}

Consider first the results in the multivariate normal case (M1) in Tables \ref{PCA_M1_k2} and \ref{PCA_M1_k3}.
 Schott's test seems to give slightly smaller size and power  estimates than the asymptotic test based on the covariance matrix. The sizes of both asymptotic tests seem to be close to the nominal value except for the small-$n$-large-$p$ cases, and the asymptotic test based on the covariance matrix
is superior to the test based on Tyler's shape matrix. The results for the bootstrap and asymptotic versions of the tests seem similar, again except for the small-$n$-large-$p$ cases
where the bootstrap tests are more conservative.  The full Tyler's estimate  suffers from the sparsity of the data more than the 3-step version. Therefore, for the models M2 and M3,
we compute the results for the much faster 3-step version only. \\

The simulation results for the noisy ICA model (M2) are reported in  Tables \ref{PCA_M2_k2} and \ref{PCA_M2_k3}.
In these cases, the asymptotic tests  work well although the estimate of unknown $\sigma_1$ is not consistent under this model assumption. The behavior of the
bootstrap tests as compared to the asymptotic tests is similar as in the previous case; only the  PCA-II strategy is valid in this case and also seems better here than
the PCA-I strategy. \\

In the model M3 the observations come from an elliptic heavy-tailed $t_5$ distribution, see the results in Tables \ref{PCA_M3_k2} and \ref{PCA_M3_k3}. Although the fourth moments are bounded, the convergence towards the limiting  distribution for the test statistic based on the covariance matrix is extremely slow  and the asymptotic tests completely fail for large $p$.
Due to outliers in the sparse data, the discrete bootstrap null distributions have the outliers as well which can be even multiplied in the bootstrap samples. Therefore
also the bootstrap tests using the covariance matrix fail for large $p$. In the spirit of robust and fast bootstrap, the 3-step Tyler's shape matrix seem to work very well and provides most reliable tests in all cases,  especially for large dimensions. Other $k$-step versions of this estimate would deserve further considerations.\\

\subsection{An example}

The standard repeated measures ANOVA  needs the assumption of spherical multivariate normality. Sphericity has then been defined
both in terms of the variances of difference scores and in terms of the variances and covariances of orthogonal contrasts to be used in the analysis, see e.g. \citet{Lane16}.
Preliminary testing for sphericity or  subsphericity is then of interest in this context.
Subsphericity indicates that there are no latent subgroups or clusters in that part of the data, and the subspherical part may then be seen simply as noise.
 To illustrate the methodology we  use some data
from the LASERI study (Cardivascular risk in young Finns study) which is available in the R package ICSNP \citep{ICSNP:2015}.
To collect these data,  223 subjects took part in a tilt-table test. For the first ten minutes the subjects were lying on a motorized table in a supine position, then the  table was tilted to a head-up position  for five minutes,  and thereafter returned   to the supine position for the last five minutes. Various hemodynamic variables were measured during the experiment. The variable considered here consists of the four measurements of the systemic vascular resistance index (SVRI)  on all subjects. The four time points were (i)  the tenth supine minute before the tilt,  the (ii) second and (iii) fifth minute during the tilt and (iv) the fifth minute in supine position after the tilting.  The 223 SVRI values at the 4 time points are shown in Figure~\ref{Fig_SVRI} (left figure).

{
\begin{figure}[thp]
\centering
\includegraphics[width=0.9\textwidth]{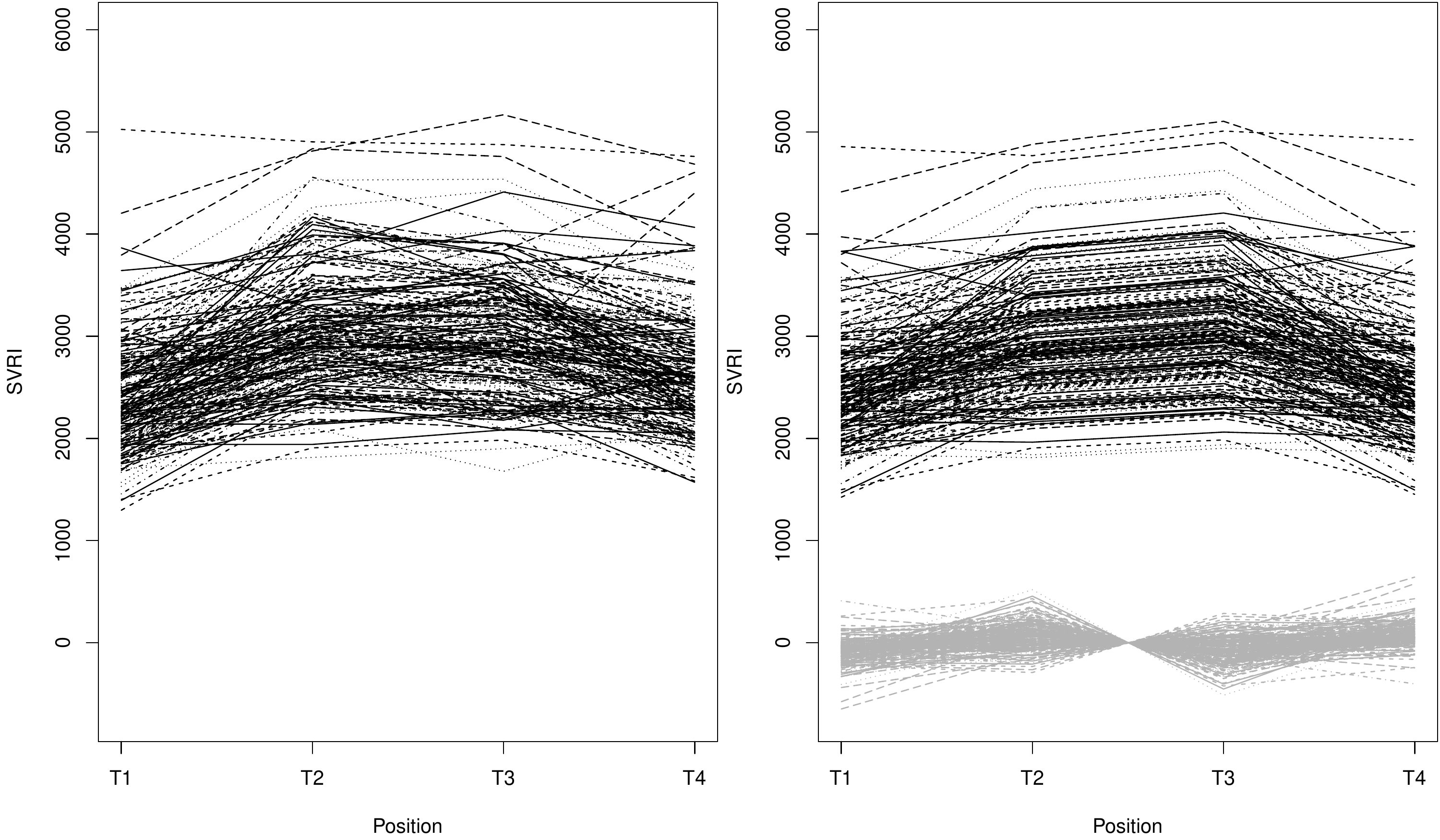}
\caption{Left figure: The original data set consisting of the SVRI values measured on  223 subjects at 4 time points.
Right figure: The estimated signal part (upper curves) and noise part  (lower part) of the same data set.}
\label{Fig_SVRI}
\end{figure}
}

We use the same scatter matrices as in the simulation study:
The eigenvalues  of the sample covariance matrix and Tyler's shape matrix   are
982935.95, 176465.68,  36213.91,  25865.65 and 8.94,  1.78,  0.30,  0.21, respectively,
and the corresponding eigenvectors are the columns of
%
\[
\left(
  \begin{array}{rrrr}
     -0.48 &  0.46 & -0.42 &  0.62 \\
     -0.51 & -0.53 & -0.56 & -0.38 \\
     -0.52 & -0.44 & 0.64  & 0.36 \\
     -0.50 &  0.56 & 0.31  & -0.59 \\
  \end{array}
\right)
\ \ \mbox{and}\ \
\left(
  \begin{array}{rrrr}
   -0.47 &  0.52 & -0.13 &  0.70 \\
   -0.51 & -0.48 & -0.70 & -0.11 \\
   -0.53 & -0.47 & 0.69  & 0.12\\
   -0.48 &  0.52 & 0.10  & -0.70 \\
  \end{array}
\right).
\]
Both scatter matrices seem to suggest that $q=2$ and that the principal components are (close) to the average and the contrast
comparing the supine and tilted positions and the two contrasts within positions.
The suggestion $q=2$ is supported by the  $p$-values for $H_{00}$, $H_{01}$ and $H_{02}$ using the two scatter matrices and three testing strategies, see Table~\ref{Tab_SVRI_pvals}.
The estimated signal and noise parts of the data using Tyler's scatter matrix are given in  Figure~\ref{Fig_SVRI} (right figure).
\begin{table}[ht]
\centering
\begin{tabular}{@{\extracolsep{4pt}}lcccccc}
  \hline
  & \multicolumn{3}{c}{Cov} & \multicolumn{3}{c}{Tyler's shape matrix}\\ \cline{2-4} \cline{5-7}
  & Asymp & PCA-I & PCA-II & Asymp & PCA-I & PCA-II \\
  \hline
$H_{00}$  & 0.000 & 0.002 & 0.002 & 0.000 & 0.002 & 0.002 \\
$H_{01}$  & 0.000 & 0.002 & 0.002 & 0.000 & 0.002 & 0.002 \\
$H_{02}$  & 0.104 & 0.130 & 0.142 & 0.064 & 0.072 & 0.064 \\ \hline
\end{tabular}
\caption{The $p$-values for $H_{00}$-$H_{02}$  using two scatter matrices and three testing strategies for the SVRI data.}
\label{Tab_SVRI_pvals}
\end{table}



\section{Testing for subspace dimension in FOBI}\label{sec:ICA}

\subsection{The model, null hypothesis and test statistic}

In the independent component (IC) model it is assumed that $\bo X=(\bo x_1,...,\bo x_n)'$ is a random sample from a distribution of
\[
\bo x=\bo A\bo z +\bo b
\]
where  $\bo A\in\mathbb{R}^{p\times p}$  is non-singular, $\bo b\in\mathbb{R}^p$,  and $\bo z$ is a random $p$-vector with
independent components standardized so that  $\mathbb{E}(\bo z)=\bo 0$ and $Cov(\bo z)=\bo I_p$.
We further assume that $\bo z=(\bo z_1',\bo z_2')'$ where the components of  $\bo z_1\in \mathbb{R}^q$ (signal) are non-Gaussian and the components of $\bo z_2\in \mathbb{R}^{p-q}$ (noise) are Gaussian.
The general idea then is to make inference on the unknown $q$, $0\le q\le p$, and to estimate the non-Gaussian signal and Gaussian noise subspaces. In this chapter we discuss some recent  tests and estimates for $q$ introduced in \citet{Nordhausen:2017} that are based on the joint use of the covariance matrix and the matrix of fourth moments.  Throughout this chapter we therefore need to assume that the fourth moments of $\bo z$ exist. \\

In the independent component analysis (ICA) it is usually assumed that $q$ is $p-1$ or $p$. If $1\le q\le p$ is allowed as in our case, the approach is
sometimes called  non-Gaussian independent component analysis (NGICA).
In the non-Gaussian component/subspace analysis (NGCA),  $\bo z_1$ and $\bo z_2$ are independent,  $\bo z_1$ is non-Gaussian and  $\bo z_2$ is Gaussian, that is, there is no $\bo a_1\in \mathbb{R}^q$ such that $\bo a_1'\bo z_1$ has a normal distribution while $\bo a_2'\bo z_2$  has a normal distribution for all $\bo a_2\in \mathbb{R}^{p-q}$. The components of $\bo z_1$ are thus allowed to be dependent in the NGCA model.
See \citet{Blanchard05}, \citet{Theis:2011} and \citet{Nordhausen:2017}. \\

In fourth order blind identification (FOBI) an  unmixing matrix $\bo W\in \mathbb{R}^{p\times p}$
and a diagonal matrix $\bo D \in \mathbb{R}^{p\times p}$ are found such that
\[
\bo W \bo S_1 \bo W'=\bo I_p
\ \ \ \ \mbox{and}\ \ \ \
\bo W \bo S_2 \bo W'=\bo D
\]
where $\bo S_1=\mathbb{E}\left[(\bo x-\mathbb{E}(\bo x) )(\bo x-\mathbb{E}(\bo x) ) '\right]$ and
$\bo S_2=\mathbb{E}\left[r^2 (\bo x-\mathbb{E}(\bo x) )(\bo x-\mathbb{E}(\bo x) ) '\right]$
with  $r^2=(\bo x-\mathbb{E}(\bo x) )' \bo S_1^{-1} (\bo x-\mathbb{E}(\bo x) )$ is the scatter matrix based on fourth moments.
The matrix $\bo W$ is called an unmixing matrix as $\bo W\bo x$ has independent components under the   assumption that $E(z_1^4),...,E(z_q^4)$ are distinct from one another and from $3$ (normal case). Write $\bo U'=\bo W\bo S_1^{1/2}$. As $\bo U'\bo U=\bo I_p$, $\bo U$ is orthogonal and $\bo W=\bo U'\bo S_1^{-1/2}$.  If
 $$\bo R:=\bo S_1^{-1/2}\bo S_2\bo S_1^{-1/2},$$ then $\bo W\bo S^{1/2} \bo R \bo S^{1/2}\bo W'= \bo U'\bo R\bo U=\bo D$ and $\bo U$ is therefore obtained from  the eigen-decomposition $\bo R=\bo U\bo D\bo U'$.
 The eigenvalue $d_i$ in $\bo D$ is then $p+2$ if and only if  $E(z_i^4)=3$, $i=1,...,p$,  and, under mild assumptions,  the eigenvalues can be used to separate the Gaussian and non-Gaussian components.
As $\bo W(F_{\bo A\bo x})\bo A\bo x$ and $\bo W(F_{\bo x})\bo x$ are the same up to sign changes, location shifts and perturbations of the coordinates and  the ordered eigenvalues of $\bo D(F_{\bo A\bo x})$ and of   $\bo D(F_{\bo x})$ are the same,  we can in our derivations assume without any loss of generality
that  $\bo A=\bo I_p$, $\bo b=\bo 0$ and   $\bo S_1=\bo I_p$, $\bo S_2=\bo R=\bo D=diag(\bo D_1,(p+2)\bo I_{p-q})$. For our approach, we also need the assumption that the
 diagonal elements in $\bo D_1$ are distinct from $p+2$.  \\

Let $\bo X=(\bo x_1,...,\bo x_n)'$ be  a random sample from the stated independent component model with $q$ non-Gaussian and $p-q$ Gaussian independent components
with an unknown dimension $q$. Write $\widehat{\bo S}_1$, $\widehat{\bo S}_2$ and $\widehat{\bo R}$  for the values of functionals $\bo S_1$, $\bo S_2$  and $\bo R$, respectively, at the empirical distribution of $\bo X$. If $\sqrt{n}(\widehat{\bo S}_1-\bo I_p)=O_P(1)$ and $\sqrt{n}(\widehat{\bo S}_2-\bo D)=O_P(1)$ then,
by Slutsky's theorem, $$\sqrt{n}(\widehat{\bo R}-\bo D)=
\sqrt{n}(\widehat{\bo S}_2-\bo D)-\frac 12 \left[ \sqrt{n}(\widehat{\bo S}_1-\bo I_p)\bo D + \bo D \sqrt{n}(\widehat{\bo S}_1-\bo I_p)  \right]+ o_P(1)$$
and the limiting multivariate normality of $\sqrt{n} vec (\widehat{\bo R}-\bo D)$ follows from the joint limiting multivariate normality of $\sqrt{n}vec(\widehat{\bo S}_1-\bo I_p, \widehat{\bo S}_2-\bo D)$ which holds if the eight moments of $\bo z$ exist.
We wish to test the null hypothesis
\[
H_{0k}:\ \ \mbox{exactly $p-k$ eigenvalues in $\bo D$ are $p+2$}
\]
stating that the dimension of the signal space is  $k$.  To test the null hypothesis $H_{0k}$, we use the test statistic
\begin{eqnarray*}
T_{k}&:=& \min_{\bo U\in \mathcal{O}^{p\times (p-k)}}  m_2\left(\bo U' (\widehat{\bo R}-(p+2)\bo I_p) \bo U\right) \\
&=&  \min_{\bo U\in \mathcal{O}^{p\times (p-k)}}  m_1\left(\bo U' (\widehat{\bo R}-(p+2)\bo I_p)^2 \bo U\right).
\end{eqnarray*}
Recall that \citet{Kankainen07} used $T_0=m_2\left(\widehat{\bo R}-(p+2)\bo I_p \right)$ to test for full multivariate normality of $\bo x$.
If
\[
\widehat{\bo U}_k=\arg \min_{\bo U\in \mathcal{O}^{p\times (p-k)}}  m_1\left(\bo U' (\widehat{\bo R}-(p+2)\bo I_p)^2 \bo U\right),
\]
then, again according to the Poincar\'{e} separation theorem, a solution of $\widehat{\bo U}_k$ is the matrix of
the eigenvectors associated with the $p-k$  eigenvalues  of $\bo R$ that are closest to $p+2$.
We can then also write
\begin{eqnarray*}
T_k &=& m_2\left(\widehat{\bo U}_k' (\widehat{\bo R}-(p+2)\bo I_p) \widehat{\bo U}_k\right) \\
&=& s^2 \left(\widehat{\bo U}_k' \widehat{\bo R} \widehat{\bo U}_k\right)+\left[m_1\left(\widehat{\bo U}_k' \widehat{\bo R} \widehat{\bo U}_k\right)-(p+2)\right]^2
\end{eqnarray*}
and  $\widehat{\bo U}_k'\widehat{\bo S}_1^{-1/2}\bo x$ is, under $H_{0k}$, an estimate for the Gaussian noise vector.


\subsection{Asymptotic tests for dimension}\label{sec:ICAas}

Consider the independent component model and, without loss of generality, presume  $\bo A=\bo I_p$ and  $\bo b=\bo 0$. Let $q$ denote the dimension of the non-Gaussian signal space, and  denote the corresponding partition by
\[
\widehat {\bo R}=\left(
   \begin{array}{cc}
    \widehat{\bo  R}_{11}  & \widehat{\bo  R}_{12} \\
     \widehat{\bo  R}_{21} & \widehat{\bo  R}_{22} \\
   \end{array}
 \right).
\]
We then have  the following result.

\begin{lemma}\label{L1:ICA}
Under the previously stated assumptions and under  $H_{0q}$,
\begin{eqnarray*}
 n  T_q &=& n\cdot  m_2\left(\widehat{\bo R}_{22}-(p+2)\bo I_{p-q}\right) +O_P(n^{-1/2}) \\
 &=& n\cdot s^2\left(\widehat{\bo R}_{22}\right)+n\left[m_1(\widehat{\bo R}_{22}) -(p+2)\right]^2 +O_P(n^{-1/2}).
\end{eqnarray*}
\end{lemma}

Note that the first term in the sum on the second row provides a test statistic for the equality of $p-q$  eigenvalues closest to $p+2$   and the second term measures the deviation
of their average from $p+2$ (Gaussian case).
Under our assumptions and under $H_{0q}$,  
these two random variables are asymptotically independent and we have the following.

\begin{theorem}\label{Th1:ICA}
Under the previously stated assumptions and under $H_{0q}$,
\[
n (p-q)T_q
\to_d  {{2\sigma_1}}  \chi^2_{\frac 12(p-q-1)(p-q+2)}+ \left({{2\sigma_1}+\sigma_2(p-q)} \right) \chi^2_1
\]
with independent chi squared variables $\chi^2_{\frac 12 (p-q-1)(p-q+2)}$ and  $\chi^2_1$  and\\
 $\sigma_1= Var\left(\|\bo z  \|^2\right)+8$ and $\sigma_2=4$.
\end{theorem}
Recall that $T_q=T_{q,1}+T_{q,2}$  where
$T_{q,1}= s^2 (\widehat{\bo U}_q' \widehat{\bo R} \widehat{\bo U}_q)$ and $T_{q,2}=[m_1\left(\widehat{\bo U}_q' \widehat{\bo R} \widehat{\bo U}_q\right)-(p+2)]^2$
provide two asymptotically independent test statistics for $H_{0q}$ as seen from the proof of the theorem.
Under the assumptions in Theorem~\ref{Th1:ICA}, $n (p-q)T_{q,1}\to_d {{2\sigma_1}}  \chi^2_{\frac 12(p-q-1)(p-q+2)}$
and  $n (p-q)T_{q,2}\to_d  \left({{2\sigma_1}+\sigma_2(p-q)} \right) \chi^2_1$.
For deriving the values of $\sigma_1$ and $\sigma_2$, see the appendix in \citet{Nordhausen:2017}. They show that the result is true even in the wider NGCA model. As seen in the proof,  $\sigma_1=AsVar((\widehat{\bo R}_{22})_{12})$
and  $\sigma_2=AsCov((\widehat{\bo R}_{22})_{11},(\widehat{\bo R}_{22})_{22} )$.  In the independent component model, we simply have $\sigma_1=\sum_{k=1}^p E(z_k^4)-p+8$
with a consistent estimate $\hat\sigma_{1a}=\frac 1n \sum_{i=1}^n\sum_{k=1}^p (\hat z_i)_k^4-p+8$
where  $\hat {\bo z}_i=\widehat{\bo W}(\bo x_i-\bar{\bo x})$, $i=1,...,n$.
In the wider NCGA model, the parameter $\sigma_1$ can be consistently estimated by
$
\hat\sigma_{1b}=\frac 1{n} \sum_{i=1}^n \|\hat {\bo z}_i \|^4-p^2+8.
$
Both estimates, $\hat\sigma_{1b}$ and  $\hat\sigma_{1b}$, are consistent in the case of the independent component model even for unknown $q$.
\\

To estimate $q$, we  consider the joint limiting behavior of test statistics $n(p-k)T_k$ for $H_{0k}$, $k=0,...,p-1$,  but under true  $H_{0q}$.
For $k=0,1,...,p-1$, write
\[
T_k^*=
m_2\left( (\bo 0,\bo I_{p-k}) (\widehat{\bo R}-(p+2)\bo I_p) (\bo 0,\bo I_{p-k})' \right).
\]
Then $T_k\le T_k^*$, $k=0,...,p-1$, and we have the following \citep{Nordhausen:2017}.

\begin{theorem}\label{Th2:ICA}
Under the previously stated assumptions and under $H_{0q}$,
\begin{itemize}
\item[(i)] for $k<q$, $T_k\to_P c_k$ for some $c_1,...,c_{q-1}>0$,
\item[(ii)] for $k=q$,
$n (p-k) T_k\to_d C_k$,
and
\item[(iii)] for $k>q$,
$  n (p-k) T_k \le n(p-k) T_k^* \to_d C_k$
\end{itemize}
where
$$ C_k\sim {2\sigma_1} \chi^2_{(p-k-1)(p-k+2)/2}+ \left({{2\sigma_1}+ {\sigma_2} (p-k)} \right) \chi^2_1$$
with independent chi squared variables $\chi^2_{(p-k-1)(p-k+2)/2}$ and  $\chi^2_1$
and $\sigma_1$ and $\sigma_2$ as in Theorem~\ref{Th2:ICA}.
\end{theorem}

As in PCA, a consistent estimate $\hat q$ of the unknown dimension  $q$ can   be based on sequential testing using the test statistics $T_k$ and corresponding critical values $c_{k,n}$, $k=0,...,p-1$,
 as suggested in the following. Other (top-down or divide and conquer) strategies again provide alternative consistent estimates.
\begin{corollary}
For all $k=0,...,p-1$,  let
  $(c_{k,n})$  be a sequence of positive real numbers such that
$c_{k,n}\to 0$ and $n{c_{k,n}}\to \infty$ as $n\to\infty$.
Then
\[
\mathbb{P}(T_k\ge c_{k,n})\ \rightarrow\
\left\{
       \begin{array}{ll}
         1, & \hbox{if $k<q$ ;} \\
         0, & \hbox{if $k\ge q$,}
       \end{array}
     \right.
\]
and
\[
\hat q=\min \{k\ :\ T_k < c_{k,n}\}\to_P q.
\]
\end{corollary}

\subsection{Bootstrap tests for dimension}

Let $q$ denote the true dimension and consider the test statistic
$T_k= m_2\left(\widehat{\bo U}_k' (\widehat{\bo R}-(p+2)\bo I_p) \widehat{\bo U}_k\right) $ for $H_{0k}$, $k=0,...,p-1$.
In the following we also need
\[
\widehat{\bo P}_k= \widehat{\bo S}_1^{1/2}\widehat{\bo U}_k\widehat{\bo U}_k' \widehat{\bo S}_1^{-1/2} \ \ \mbox{and}\ \
\widehat{\bo Q}_k=\bo I_p-\widehat{\bo P}_k,
\]
which are the estimated projection matrices  (with respect to Mahalanobis inner product) to the noise and signal subspaces, respectively.

 To obtain the $p$-value for $T_k$, the bootstrap samples are generated, as in PCA,  from a distribution for which  the null hypothesis $H_{0k}$ is true under the stated model (even if $k\ne q$) and  which is as similar as possible to the empirical distribution of $\bo X$. We suggest again  two procedures. The first one  is for testing the hypothesis $H_{0k}$  in the IC model  and the second one in the wider NGCA model, see \citet{Nordhausen:2017}. The bootstrap $p$-values are obtained as in PCA with $M$ bootstrap samples.\\

{\bf  Bootstrap strategy FOBI-I (IC model):}\ \
{\it
\begin{enumerate}
\item Start with centered $\bo X\in \mathbb{R}^{n\times p}$ and compute $\bar{\bo x}$ and $\widehat{\bo W}=(\widehat{\bo W}_1',\widehat{\bo W}_2')'$ where
$\widehat{\bo W}_2=\widehat{\bo U}_k'\widehat{\bo S}_1^{-1/2}$.
\item  Write  $\widehat{\bo Z}=(\bo X-\bo 1_n\bar{\bo x}')\widehat{\bo W}'$ and further
$\widehat{\bo Z}=(\widehat{\bo Z}_1,\widehat{\bo Z}_2)$ where $\widehat{\bo Z}_2\in \mathbb{R}^{n\times (p-k)}$.
\item Let ${\bo Z}_1^*\in \mathbb{R}^{n\times k}$ for a matrix of independent componentwise bootstrap samples of size $n$ from  $\widehat{\bo Z}_1 $.
\item Let $\bo Z_2^*\in \mathbb{R}^{n\times (p-k)}$ be a random sample of size $n$ from $N_{p-k}(\bo 0,\bo I_{p-k})$.
\item Write  ${\bo Z}^*=(\bo Z_1^*, \bo Z_2^*)$.
\item Write $\bo X^*={\bo Z}^* (\widehat{\bo W}')^{-1}+\bo 1_n\bar{\bo x}'$.
\end{enumerate}}

{\bf  Bootstrap strategy FOBI-II (NGCA model):}\ \ {\it
\begin{enumerate}
\item Starting with $\bo X\in \mathbb{R}^{n\times p}$, compute $\bar{\bo x}, \widehat{\bo S}_1$, $\widehat{\bo S}_2$, $\widehat{\bo R}$,
$\widehat{\bo U}_k$, $\widehat{\bo P}_k$  and $\widehat{\bo Q}_k$.
 \item Take a bootstrap sample   $\widetilde{\bo X}=(\tilde{\bo x}_1, \ldots, \tilde{\bo x}_n)'$ of size $n$ from  $\bo X$.
\item For the noise space to be Gaussian, transform
\[ \bo x_i^*= [\widehat{\bo Q}_k(\tilde{\bo x}_i-\bar{\bo x}) +\widehat{\bo S}_1^{1/2} \widehat{\bo U}_k \bo o_i]+\bar{\bo x} ,\ \ i=1,...,n,  \]
where  $\bo o_1, \ldots, \bo o_n$ are iid  from $N_{p-k}(\bo 0,\bo I_{p-k})$.
 \item $\bo X^* =(\bo x_1^*,...,\bo x_n^*)'$.
\end{enumerate}}

In the case of the  FOBI-I strategy, the bootstrap null distribution $F_{k,n}(\bo x)$ is the average
\[
\frac 1{n^k}
\sum_{i_1,...,i_k=1}^n
\mathbb{E}_{\bo o_{i_1...i_k}}\left[
\mathbb{I}\left(
\widehat {\bo W}^{-1} \left(
                        \begin{array}{c}
                          (\bo e_{i_1},...,\bo e_{i_k})'(\bo X-\bo 1_n\bar{\bo x}')\widehat{\bo W}_1' \\
                          \bo o_{i_1...i_k} \\
                        \end{array}
                      \right)
  +\bar{\bo x}\le x \right) \right]
\]
where the $\bo o_{i_1...i_k}$'s are from $N_{p-k}(\bo 0,\bo I_{p-k})$ and the $\bo e_i$'s (with the $i$th element one and other elements zero)
 are in $\mathbb{R}^n$, and
in the FOBI-II strategy, the bootstrap samples for $H_{0k}$ are generated from the distribution $F_{k,n}(\bo x)$ that is the average
\[
\frac 1n \sum_{i=1}^n \mathbb{E}_{o_i} \left[ \mathbb{I} \left([\widehat{\bo Q}_k({\bo x}_i-\bar{\bo x}) +\widehat{\bo S}_1^{1/2} \widehat{\bo U}_k \bo o_i]+\bar{\bo x}\le \bo x
\right)\right]
\]
where  $\bo o_1, \ldots, \bo o_n\sim N_{p-k}(\bo 0,\bo I_{p-k})$.\\

As in the  PCA bootstrap asymptotics, let $\bo X_N^*$ be a random sample of size $N$ from $F_{n,k}$. As these observations come from  the ICA and NGCA models, respectively, with true $H_{0k}$ and known (data based) parameters $\sigma_1=\hat\sigma_{1a}$ or $\sigma_1=\hat\sigma_{1b}$ and $\sigma_2=4$, the limiting (conditional and unconditional) distribution of $NT_k(\bo X_N^*)$ is as given in Theorem~\ref{Th1:ICA}. For large $n$, the limiting distribution then provides the approximation for $nT_k(\bo X^*)$ as well.

\subsection{A simulation study}

To compare the asymptotic and bootstrap tests, we consider the  independent component models  where $\bo z_1,...,\bo z_n\in \mathbb{R}^p$ have the following marginal signal distributions:
\begin{description}
  \item[M1:] exponential, $\chi^2_2$, uniform, normal, ..., normal
  \item[M2:] exponential, $\chi^2_2$, $t_5$, normal, ..., normal
\end{description}
Hence,  $q=3$ in both cases and we use the dimensions $p=6$ (3 Gaussian components) and $p=15$ (12 Gaussian components).
The only difference between the models M1 and M2 is that, in the model M2 the uniform distribution (low kurtosis) is replaced by the $t_5$ distribution (high kurtosis).
As the tests only use the eigenvalues of $\widehat {\bo D}$, the simulation results are the same for any choices of  $\bo A$ and $\bo b$.
For the simulations in the case of the NCGA model, see \citet{Nordhausen:2017}. \\

We compare the four tests with $p$-values obtained from (i) the asymptotic null distribution using $\hat \sigma_{1a}$ (Asy1) ,  (ii) the asymptotic null distribution using $\hat \sigma_{1b}$ (Asy2),
(iii) the bootstrap null distribution using the strategy FOBI-I (Boot1), and (iv) the bootstrap null distribution using the strategy FOBI-II (Boot2) . Note that Asy1 and Boot1 assume the IC model while Asy2 and Boot2
are valid in the wider NGCA model.\\

For all samples $\bo X\in \mathbb{R}^{n\times p}$ generated from models M1 and M2, the $p$-values based on the asymptotic null distribution (Asy1, Asy2) and the bootstrap $p$-values based on  $M=200$ bootstrap samples (Boot1, Boot2) were computed, and the sampling of $\bo X$ and computation of $p$-values was repeated 2000 times.
The null hypothesis was rejected if the observed $p$-value is smaller than  0.05.
Tables~\ref{Tab_ICA_M1_ICA}-\ref{Tab_ICA_M2_NGCA} report the average proportions  of rejections for $H_{02}$, $H_{03}$ (true) and $H_{04}$
in 2000 repetitions. All simulations were also repeated using $M=500$. The changes in the results were negligible and therefore not reported here.
\begin{table}[ht]
\centering
\begin{tabular}{@{\extracolsep{4pt}} rrrrrrrr@{}}
  \hline
 \multicolumn{2}{c}{} & \multicolumn{2}{c}{$H_{02}$} & \multicolumn{2}{c}{$H_{03}$ (true)} & \multicolumn{2}{c}{$H_{04}$}\\ \cline{3-4} \cline{5-6}  \cline{7-8}
 p & n & Asy1 & Boot1 & Asy1 & Boot1 & Asy1 & Boot1 \\
  \hline
   6 & 200   & 0.086 & 0.404  &  0.006 & 0.107 &0.000 & 0.066 \\
    & 500   & 0.687 & 0.845  &  0.026 & 0.068 &0.001 & 0.039 \\
    & 1000  & 0.994 & 0.996  &  0.044 & 0.062 &0.002 & 0.038 \\
    & 2000  & 1.000 & 1.000  &  0.052 & 0.063 &0.004 & 0.038 \\
    & 5000  & 1.000 & 1.000  &  0.061 & 0.066 &0.003 & 0.041 \\
    & 10000 & 1.000 & 1.000  &  0.050 & 0.051 &0.005 & 0.043 \\  \hline
  15 & 200   & 0.002 & 0.073  &  0.000 & 0.058 &0.000 & 0.045 \\
   & 500   & 0.033 & 0.129  &  0.002 & 0.096 &0.000 & 0.067 \\
   & 1000  & 0.125 & 0.210  &  0.016 & 0.121 &0.000 & 0.067 \\
   & 2000  & 0.502 & 0.562  &  0.044 & 0.096 &0.002 & 0.055 \\
   & 5000  & 0.997 & 0.999  &  0.071 & 0.079 &0.005 & 0.058 \\
   & 10000 & 1.000 & 1.000  &  0.073 & 0.077 &0.007 & 0.062 \\
   \hline
\end{tabular}
\caption{
Rejection rates for $H_{02}$-$H_{04}$ in  the model M1 with two dimensions $p=6,15$ and sample sizes $n=200,...,10000$.
The tests Asy1 and Boot1 assume the independent component model. }
\label{Tab_ICA_M1_ICA}
\end{table}

\begin{table}
\centering
\begin{tabular}{@{\extracolsep{4pt}} rrrrrrrr@{}}
  \hline
 \multicolumn{2}{c}{} & \multicolumn{2}{c}{$H_{02}$} & \multicolumn{2}{c}{$H_{03}$ (true)} & \multicolumn{2}{c}{$H_{04}$}\\ \cline{3-4} \cline{5-6}  \cline{7-8}
 p & n & Asy2 & Boot2 & Asy2 & Boot2 & Asy2 & Boot2 \\
  \hline
   6 & 200   &  0.128 & 0.333  & 0.008 & 0.065 & 0.000 & 0.017  \\
    & 500   &  0.723 & 0.800  & 0.020 & 0.052 & 0.001 & 0.013  \\
    & 1000  &  0.996 & 0.996  & 0.037 & 0.053 & 0.001 & 0.014  \\
    & 2000  &  1.000 & 1.000  & 0.041 & 0.049 & 0.004 & 0.012  \\
    & 5000  &  1.000 & 1.000  & 0.052 & 0.055 & 0.002 & 0.011  \\
    & 10000 &  1.000 & 1.000  & 0.039 & 0.042 & 0.003 & 0.013  \\  \hline
  15 & 200   &  0.001 & 0.043  & 0.000 & 0.024 & 0.000 & 0.012  \\
   & 500   &  0.038 & 0.074  & 0.002 & 0.046 & 0.000 & 0.024  \\
   & 1000  &  0.154 & 0.141  & 0.012 & 0.053 & 0.000 & 0.018  \\
   & 2000  &  0.533 & 0.471  & 0.032 & 0.057 & 0.001 & 0.016  \\
   & 5000  &  0.998 & 0.997  & 0.049 & 0.059 & 0.002 & 0.019  \\
   & 10000 &  1.000 & 1.000  & 0.045 & 0.051 & 0.004 & 0.014  \\
   \hline
\end{tabular}
\caption
{Rejection rates for $H_{02}$-$H_{04}$ in  the model M1 with two dimensions $p=6,15$ and sample sizes $n=200,...,10000$.
The tests Asy2 and Boot2 are valid in the wider NGCA model. }
\label{Tab_ICA_M1_NGCA}
\end{table}

\begin{table}
\centering
\begin{tabular}{@{\extracolsep{4pt}} rrrrrrrr@{}}
  \hline
 \multicolumn{2}{c}{} & \multicolumn{2}{c}{$H_{02}$} & \multicolumn{2}{c}{$H_{03}$ (true)} & \multicolumn{2}{c}{$H_{04}$}\\ \cline{3-4} \cline{5-6}  \cline{7-8}
 p & n & Asy1 & Boot1 & Asy1 & Boot1 & Asy1 & Boot1  \\
  \hline
   6 & 200   &  0.326 & 0.493  & 0.006 & 0.120 &  0.000 & 0.073 \\
    & 500   &  0.808 & 0.861  & 0.028 & 0.093 &  0.002 & 0.059 \\
    & 1000  &  0.980 & 0.987  & 0.043 & 0.078 &  0.003 & 0.043 \\
    & 2000  &  1.000 & 1.000  & 0.043 & 0.055 &  0.002 & 0.038 \\
    & 5000  &  1.000 & 1.000  & 0.045 & 0.055 &  0.002 & 0.034 \\
    & 10000 &  1.000 & 1.000  & 0.049 & 0.053 &  0.002 & 0.039 \\  \hline
  15 & 200   &  0.038 & 0.203  & 0.000 & 0.082 &  0.000 & 0.060 \\
   & 500   &  0.371 & 0.534  & 0.005 & 0.103 &  0.000 & 0.065 \\
   & 1000  &  0.777 & 0.830  & 0.018 & 0.084 &  0.001 & 0.065 \\
   & 2000  &  0.986 & 0.989  & 0.030 & 0.059 &  0.001 & 0.046 \\
   & 5000  &  1.000 & 1.000  & 0.035 & 0.050 &  0.004 & 0.046 \\
   & 10000 &  1.000 & 1.000  & 0.040 & 0.050 &  0.002 & 0.034 \\
   \hline
\end{tabular}
\caption
{Rejection rates for $H_{02}$-$H_{04}$ in  the model M2 with two dimensions $p=6,15$ and sample sizes $n=200,...,10000$.
The tests Asy1 and Boot1 assume the independent component model. }
\label{Tab_ICA_M2_ICA}
\end{table}

\begin{table}
\centering
\begin{tabular}{@{\extracolsep{4pt}} rrrrrrrr@{}}
  \hline
 \multicolumn{2}{c}{} & \multicolumn{2}{c}{$H_{02}$} & \multicolumn{2}{c}{$H_{03}$ (true)} & \multicolumn{2}{c}{$H_{04}$}\\ \cline{3-4} \cline{5-6}  \cline{7-8}
 p & n & Asy2 & Boot2 & Asy2 & Boot2 & Asy2 & Boot2 \\
  \hline
   6 & 200   &  0.310 & 0.470  & 0.005 & 0.051  & 0.000 &0.018   \\
    & 500   &  0.796 & 0.853  & 0.022 & 0.057  & 0.002 &0.019   \\
    & 1000  &  0.979 & 0.985  & 0.033 & 0.057  & 0.001 &0.013   \\
    & 2000  &  0.999 & 1.000  & 0.035 & 0.049  & 0.002 &0.012   \\
    & 5000  &  1.000 & 1.000  & 0.042 & 0.044  & 0.002 &0.007   \\
    & 10000 &  1.000 & 1.000  & 0.045 & 0.049  & 0.002 &0.008   \\  \hline
  15 & 200   &  0.033 & 0.157  & 0.001 & 0.031  & 0.000 &0.018   \\
   & 500   &  0.363 & 0.502  & 0.002 & 0.035  & 0.000 &0.012   \\
   & 1000  &  0.762 & 0.810  & 0.011 & 0.044  & 0.001 &0.015   \\
   & 2000  &  0.985 & 0.987  & 0.018 & 0.038  & 0.000 &0.008   \\
   & 5000  &  1.000 & 1.000  & 0.029 & 0.035  & 0.003 &0.013   \\
   & 10000 &  1.000 & 1.000  & 0.035 & 0.039  & 0.001 &0.009   \\
   \hline
\end{tabular}
\caption
{Rejection rates for $H_{02}$-$H_{04}$ in  the model M2 with two dimensions $p=6,15$ and sample sizes $n=200,...,10000$.
The tests Asy2 and Boot2 are valid in the wider NGCA. }
\label{Tab_ICA_M2_NGCA}
\end{table}

In the following we comment on the simulation results in Tables~\ref{Tab_ICA_M1_ICA}--\ref{Tab_ICA_M2_NGCA}.
 For the true null hypothesis, $p=6$ and $n\ge 1000$, all rejection rates are close to the target size value 0.05.
 For large $p$ and small $n$, the null rejection rates of the bootstrap tests are much closer to the target value 0.05 than the rejection rates of the asymptotic tests. This is due to the fact that the asymptotic tests neglects the variation coming form the estimation of the subspace 
 which is hard in this case.
 Theorem~\ref{Th2:ICA}(iii) implies that, for $k>q$, the $p$-values obtained from the asymptotic null distribution tend to be  large and the rejection rates are then expected to be smaller than the target value 0.05. The same seems to be true for the bootstrap tests although  Boot1 yields
rejection rates which are quite close to 0.05. 
  For $k<q$, the powers naturally increase with $n$ and decrease with $p$.
For simulations for the NGCA model, see also \citet{Nordhausen:2017}.
To conclude, if one does not trust in the IC model, it is safe and valid to use Asy2 and Boot2 that are valid also in the wider MGCA model.

\subsection{An example}

ICA is often illustrated using mixed images. Following this tradition, we mix 6 grey scale images: Two of the images are the pictures of a cat and a forest road, available in the R package ICS \citep{Nordhausen_b:2008},  and the remaining four images are just Gaussian noise. The images have $130\times 130$ pixels and the six original images can be presented as a matrix $\bo Z\in \mathbb{R}^{n\times p}$
 with  $n=16900$ pixels and $p=6$ rows.  The observed mixed images are then $\bo X=\bo Z\bo A'+\bo 1_n\bo b'$ and the idea  is to recover the two (signal) images.
Note that the rows of $\bo X$ are not independent in this example but FOBI uses the marginal distribution of the column elements rather than their joint distribution.

\begin{figure}[thp]
\centering
\includegraphics[width=0.9\textwidth]{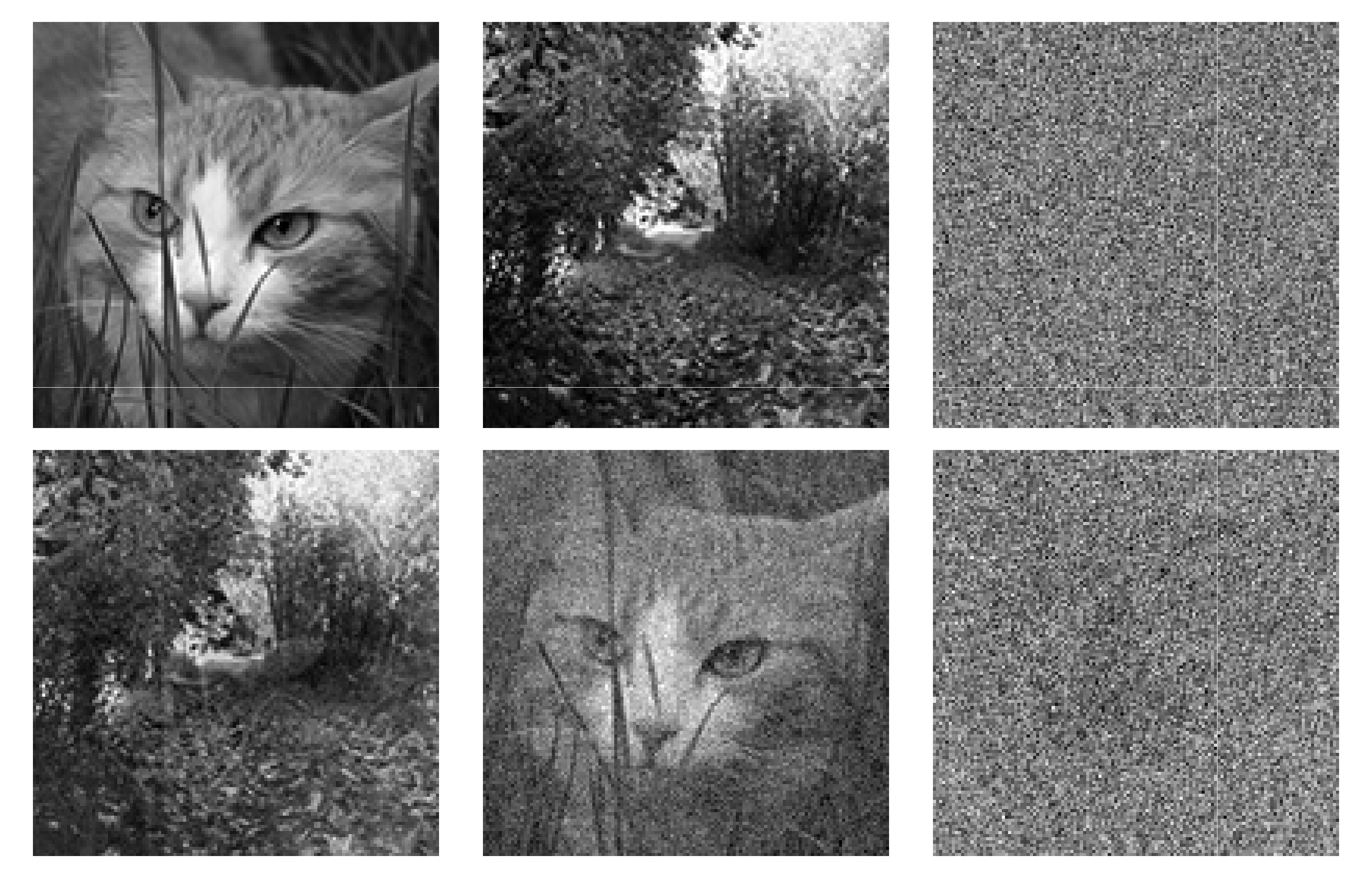}
\caption{The first three images in $\bo Z$ (upper row) and in $\widehat{\bo Z}$ (lower row).}
\label{Fig_Pics}
\end{figure}

The first three columns of the $\bo Z$ and $\widehat {\bo  Z}=\bo X \widehat {\bo W} '$ are given on the first and second row of Figure~\ref{Fig_Pics}, respectively. Note that the result on the second row would be the same for any choices of $\bo A$ and $\bo b$. The ordered eigenvalues (with respect to the squared deviation from $p+2=8$)  of $\widehat{\bo R}$ are $9.00, 8.27, 7.92, 8.04, 7.97$ and  $8.00$. The $p$-values for $H_{01}$-$H_{04}$ both all the tests are given Table~\ref{Tab_Mix_pvals}. Note that the boostrap tests here use $m=500$ bootstrap samples. In this examples all four tests nicely agree and the false hypothesis $H_{01}$ is rejected and the true hypothesis $H_{02}$ is the first to be accepted at level $\alpha=0.05$.

\begin{table}
\centering
\begin{tabular}{@{\extracolsep{4pt}} rrrrr@{}}
  \hline
 \multicolumn{1}{c}{} & \multicolumn{2}{c}{ICA} & \multicolumn{2}{c}{NGCA} \\ \cline{2-3} \cline{4-5}
 & Asymp & Boot & Asymp & Boot  \\
  \hline
  \hline
$H_{0,1}$ &  0.000 & 0.002 & 0.000 & 0.002 \\
$H_{0,2}$ &   0.211 & 0.082 & 0.206 & 0.116 \\
$H_{0,3}$ &   0.878 & 0.940 & 0.873 & 0.880 \\
$H_{0,4}$ &   0.810 & 0.778 & 0.806 & 0.729 \\
   \hline
\end{tabular}
\caption{The $p$-values for $H_{01}$-$H_{04}$ for the image data.}
\label{Tab_Mix_pvals}
\end{table}



\section{Testing for subspace dimension in SIR}\label{sec:SIR}

\subsection{The model, null hypothesis and test statistic}

In this chapter we assume that
$$
(\bo y,\bo X)
 =\left(\left(
          \begin{array}{c}
            y_1 \\
            \bo x_1 \\
          \end{array}
        \right), ...,\left(
                       \begin{array}{c}
                         y_n \\
                         \bo x_n \\
                       \end{array}
                     \right)\right)'\in \mathbb{R}^{n\times (p+1)}
$$ is a random sample from a distribution of $(y,\bo x')'$ where
\[
\bo x=\bo A \bo z+\bo b,
\]
 $\bo A\in\mathbb{R}^{p\times p}$ is non-singular, $\bo b\in\mathbb{R}^p$ and $\bo z=(\bo z_i',\bo z_2')'$ is a random $p$-vector with $\mathbb{E}(\bo z)=\bo 0$, $Cov(\bo z)=\bo I_p$
and  $(y,\bo z_1')' \independent  \bo z_2$.
  If $\bo z_1\in \mathbb{R}^q$ and  $\bo z_2\in\mathbb{R}^{p-q}$, with q being the smallest value for which this condition holds, then they correspond respectively to the  signal and noise parts of $\bo z$.The partition  $\bo z=(\bo z_i',\bo z_2')'$ is then unique up to transformations $\bo z_1\to \bo O_1\bo z_1$ and  $\bo z_2\to \bo O_2\bo z_2$ with  $\bo O_1\in \mathcal{O}^{q\times p}$
and $\bo O_2\in \mathcal{O}^{(p-q)\times (p-q)}$. The aim is again to test and estimate the unknown dimension $q$ and then find the  projections to the well defined signal and noise subspaces of $\bo x$.

\begin{remark}
Note that our  assumption $(y,\bo z_1')' \independent  \bo z_2$  is stronger than the regular assumptions in sliced inverse regression and related methods: In classical SIR  and SAVE approaches the dependence conditions are for example (i)  $y\independent  \bo z_2| \bo z_1$ and
$E(\bo z_2|\bo z_1)=\bo 0\ a.s.$  (linearity condition) for SIR  and (ii)  $y\independent  \bo z_2| \bo z_1$, $E(\bo z_2|\bo z_1)=\bo 0$ and  $Cov(\bo z_2|\bo z_1)=\bo I_{p-q}\ a.s.$ for SAVE.  Alternative or additional  assumptions needed for easy and tractable asymptotics have been given in the literature such as the assumption that $\bo z$ is multivariate normal \citep{Li91}
or that the conditional covariance $Cov(\bo z|y)$  is constant \citep{BuraCook01}. See Section~\ref{ssec:SIRas} for more discussion.
Under our strong assumption, bootstrap samples from a true null distributions are easily generated as shown in Section~\ref{ssec:SIRboot}.
\end{remark}

In the sliced inverse regression (SIR) one finds a  transformation matrix $\bo W\in \mathbb{R}^{p\times p}$
and a diagonal matrix $\bo D\in \mathbb{R}^{p\times p}$  such that
\[
\bo W \bo S_1 \bo W'=\bo I_p
\ \ \ \ \mbox{and}\ \ \ \
\bo W \bo S_2 \bo W'=\bo D
\]
with $\bo S_1:=\mathbb{E}\left[(\bo x-\mathbb{E}(\bo x)  )(\bo x-\mathbb{E}(\bo x)  ) '\right]$  and
$\bo S_2:=\mathbb{E}\left[ \mathbb{E}(\bo x-\mathbb{E}(\bo x)  |y) \mathbb{E}(\bo x-\mathbb{E}(\bo x)  |y)'\right]$.
Under our assumptions, the diagonal elements  in $\bo D$ are
\[
d_1\ge ... \ge d_q \ge d_{q+1}=  ... = d_p=0.
\]
Again, as in ICA,  $\bo W=\bo U'\bo S_1^{-1/2}$ with some orthogonal $\bo U\in \mathbb{R}^{p\times p}$ and, if $\bo R:=\bo S_1^{-1/2}\bo S_2\bo S_1^{-1/2}$ then
$\bo U$ is the matrix of eigenvectors of  $\bo R$.\\

In practice, the random variable $y$ is  replaced by its discrete approximation as follows. Let $\mathbb{S}_1,\ldots,\mathbb{S}_H$ be H disjoint intervals (slices) such that $\mathbb{R}=\mathbb{S}_1+\ldots+\mathbb{S}_H$ and let $y^{d}:=\sum_{h=1}^H y_h  \mathbb{I}(y\in \mathbb{S}_h)$ for some choices $y_h\in \mathbb{S}_h$, $h=1,\ldots,H$, independent of $\bo z$.  The random variable $y^{d}$ can then be seen as a discrete approximation of continuous random variable $y$.  ($\mathbb{I}(y\in \mathbb{S}_h)=1$ if $y\in \mathbb{S}_h$
 and zero otherwise.) Then naturally also $(y^d,\bo z_1')' \independent  \bo z_2$. The sliced inverse regression (SIR)
then just refers to the use of the inverse regression $\mathbb{E} (\bo x-\mathbb{E}(\bo x) |y^{d})$ and the corresponding supervised scatter matrix
$$\bo S_2=\mathbb{E}\left[ \mathbb{E}(\bo x-\mathbb{E}(\bo x) |y^d) \mathbb{E}(\bo x-\mathbb{E}(\bo x) |y^d)'\right]$$ in the analysis of the data.
With this choice of $\bo S_2$, we still have $d_1\ge ... \ge d_q \ge d_{q+1}=  ... = d_p=0$.
Next write $\bo\mu:=\mathbb{E}(\bo x)$ and $\bo\Sigma:=Cov(\bo x)$,  and  $\bo\mu_h:=\mathbb{E}(\bo x| y\in \mathbb{S}_h)$, $\bo\Sigma_h:=Cov(\bo x | y\in \mathbb{S}_h )$ and $p_h=\mathbb{P} (y\in \mathbb{S}_h)$,
$h=1,...,H$.  Then
\[
\bo S_1=\bo\Sigma \ \ \mbox{and}\ \ \bo S_2=\sum_{h=1}^H p_h (\bo\mu_h-\bo \mu) (\bo\mu_h-\bo \mu)'.
\]

Consider next the corresponding sample statistics.
  For the estimates of $\bo S_1$ and $\bo S_2$,  write
\[
\widehat{\bo S}_1=\frac 1n \sum_{i=1}^n (\bo x_i-\bar{\bo x})(\bo x_i-\bar{\bo x})'
\ \ \mbox{and} \ \
\widehat{\bo S}_2=\frac 1n \sum_{h=1}^H n_h (\bar{\bo x}_h-\bar{\bo x})(\bar{\bo x}_h-\bar{\bo x})'
\]
where $\bar{\bo x}_h= \frac 1{n_h}\sum_{i=1}^n \bo x_i \mathbb{I}(y_i\in \mathbb{S}_h)$ and  $n_h=\sum_{i=1}^n  \mathbb{I}(y_i\in \mathbb{S}_h)$, $ h=1,...,H$.
Note that $np\cdot m_1 (\widehat{\bo S}_1^{-1}\widehat{\bo S}_2)$ is the well-known Pillai's trace statistic for testing $H_0:\ \bo\mu_1=...=\bo \mu_H$ under the assumption that
$\bo\Sigma_1=...=\bo\Sigma_H$ with the limiting null distribution $\chi^2_{(H-1)p}$.

Furthermore, let $\widehat{\bo R}=\widehat{\bo S}_1^{-1/2} \widehat{\bo S}_2 \widehat{\bo S}_1^{-1/2}$.
We wish to test the null hypothesis
\[
H_{0k}:\ \ d_1\ge ... \ge d_{k}>d_{k+1}=...=d_p=0
\]
stating that the dimension of the signal space is exactly  $k$.
To test the null hypothesis, we use a  natural test statistic that is the average of the $p-k$ smallest eigenvalues of $\widehat{\bo R}$, that is,
\[
T_{k}:=  m_1(\widehat{\bo U}_k'\widehat{\bo R} \widehat{\bo U}_k)
\]
where the columns of $\widehat{\bo U}_k\in \mathcal{O}^{p\times(p-k)}$ are the eigenvectors corresponding the smallest $p-k$ eigenvalues of $\widehat{\bo R}$.

\subsection{Asymptotic tests for dimension}\label{ssec:SIRas}

As the eigenvalues of $\widehat{\bo R}$ are invariant under affine transformations,  we can assume without loss of generality that $(\bo y,\bo X)$ is a random
sample from a SIR model with $\bo A=\bo I_p$ and $\bo b=\bo 0$. This implies $\bo S_1 = \bo I_p$ and $\bo \mu = \bo 0$.
We assume that the number of slices $H>q+1$,  the slices  $\mathbb{S}_1,\ldots,\mathbb{S}_H$  do not change with $n$, and the related
$\bo S_2 = \bo R = \bo D=diag(\bo D_1,\bo 0)$ with a full-rank $\bo D_1\in \mathbb{R}^{q\times q}$. The assumption thus states that, with selected $H$ slices and
by using SIR, one can find the full $q$-dimensional signal space.

Let $f_h = n_h/n$, $h=1,...,H$, and write
\[
\widehat{\bo B} = \widehat{\bo S}_1^{-1/2} \left(\sqrt{f_1} (\bar{\bo x}_1 -\bar{\bo x}), ~ \cdots, ~ \sqrt{f_H}(\bar{\bo x}_H -\bar{\bo x}) \right).
\]
Then  $\widehat{\bo R} = \widehat{\bo B}\widehat{\bo B}^\prime$ and, with $\bo\pi=(\sqrt{p}_1,...,\sqrt{p}_H)'$,
\[
\widehat{\bo B}\to {\bo B}:= \left({\bo \mu}_1, \cdots, {\bo \mu}_H  \right)diag(\bo \pi) =
 \left( \begin{array}{cc} {\bo D}_1^{1/2}  & {\bo 0}  \\ {\bo 0} & {\bo 0}  \end{array} \right) \bo Q
\]
for some $\bo Q\in \mathcal{O}^{H\times H}$ where $\bo Q=(\bo Q_1',\bo Q_2')'$ and $\bo Q_1\in \mathcal{O}^{q\times H}$ satisfies $\bo Q_1\bo\pi=0$.
With the correct $\bo Q$ and correct dimension $q$, we have the partitions
\[
\widehat{\bo B}=\left(
                  \begin{array}{c}
                    \widehat{\bo B}_1 \\
                    \widehat{\bo B}_2 \\
                  \end{array}
                \right)
\ \ \mbox{and}\ \
\widehat{\bo B}\bo Q'
=
\left(
  \begin{array}{cc}
   \widehat{\bo B}_1 \bo Q_1'  & \widehat{\bo B}_1 \bo Q_2' \\
   \widehat{\bo B}_2 \bo Q_1'  & \widehat{\bo B}_2 \bo Q_2' \\
  \end{array}
\right).
\]
An asymptotic approximation to the distribution of $T_q=m_1(\widehat{\bo U}_q' \widehat{\bo R}\widehat{\bo U}_q)$
can now be stated as follows.

\begin{lemma}\label{L1:SIR}
Under the previously stated assumptions and under  $H_{0q}$,
\[ n \cdot T_q = n \cdot m_1(\widehat{\bo B}_2 \bo Q_2'\bo Q_2 \widehat{\bo B}_2') + O_P(n^{-1/4}).
\]
\end{lemma}
Note that, in this setting,  with $\bo U_q'=\left(\bo 0,\bo I_{p-q}\right)$,
\[ \bo U_q' \widehat{\bo R} \bo U_q  = \widehat{\bo B}_2 \widehat{\bo B}_2'=\widehat{\bo B}_2 \bo Q_1'\bo Q_1 \widehat{\bo B}_2'+\widehat{\bo B}_2 \bo Q_2'\bo Q_2 \widehat{\bo B}_2'.\]
Consequently, unlike in Lemmas \ref{L1:PCA} and \ref{L1:ICA} for PCA and ICA asymptotics, the asymptotic approximation given in Lemma~\ref{L1:SIR} is not obtained by simply replacing
$\widehat{\bo U}_q$ by ${\bo U}_q$ within the definition of $T_q$.
The limiting distribution of $n (p-q) T_q$ is then  given in the following.

\begin{theorem}\label{Th1:SIR}
Under our assumptions and under $H_{0q}$,
 $ n (p-q) T_q  \to_d\chi^2_{(p-q)(H-q-1)}.$
\end{theorem}

The same limiting distribution is given in Theorem 5.1 in \citet{Li91} and  in Corollary 1 in \citet{BuraCook01} under the conditional
independence  $y\independent  \bo z_2| \bo z_1$ and under the linearity condition $E(\bo z_2|\bo z_1)=\bo 0\ a.s.$. In the former,
the theorem is stated under an additional assumption that the distribution of $\bo z$ is multivariate normal, but within the proof
it is noted that it in fact holds if $Cov(\bo z_2|y)$ does not depend on $y$. In the latter, the above theorem is stated
under the additional assumption that $Cov(\bo z|y)$ does not depend on $y$, but from their proof it can be noted that they only need
this to hold for $Cov(\bo z_2|y)$. In our setting, this condition obviously holds since $\bo z_2 \independent y$. For
completeness, a proof to Theorem \ref{Th1:SIR} is given in the Appendix. Note that for $q \ge H-1$, $T_q = 0$.

To estimate $q$, we  consider the  limiting behavior of test statistics $n(p-k)T_k$ for $H_{0k}$, $k=0,...,H-1$,  when in fact $H_{0q}$ is true.
We write
\[
T_k^*=m_1((\bo I_{p-k},\bo 0)\widehat{\bo U}_q' \widehat{\bo R}\widehat{\bo U}_q(\bo I_{p-k},\bo 0)'), \ \ k=q+1,...,H-1
\]
and then have the following.

\begin{theorem}\label{Th2:ICA}
Under the previously stated assumptions and under $H_{0q}$,
\begin{itemize}
\item[(i)] for $k<q$, $T_k\to_P c_k$ for some $c_1,...,c_{q-1}>0$,
\item[(ii)] for $k=q$,
$n (p-k) T_k\to_d \chi^2_{(p-q)(H-q-1)} $,
and
\item[(iii)] for $k>q$, $\mathbb{P}(T_k\le T_k^*)\to 1$ and
$ n(p-k)T_k^* \to_d \chi^2_{(p-k)(H-q-1)} $
\end{itemize}
\end{theorem}

As in PCA and ICA, a consistent estimate $\hat q$ of the unknown dimension  $q$ can found with the bottom-up sequential testing strategy as follows.
Again alternative testing strategies may be used to find computationally faster and consistent estimates.

\begin{corollary}
For all $k=0,...,H-1$,  let
  $(c_{k,n})$  be a sequence of positive real numbers such that
$c_{k,n}\to 0$ and $n{c_{k,n}}\to \infty$ as $n\to\infty$.
Then
$
\hat q=\min \{k\ :\ T_k < c_{k,n}\}\to_P q.
$
\end{corollary}

\subsection{A bootstrap test for dimension}\label{ssec:SIRboot}

We consider the hypotheses
$H_{0k}$ saying that the rank of $\bo D$  is $k$, $k=1,...,H-1$.
Bootstrap samples are then to be generated from a null distribution for which $(y,\bo z_1')'\independent \bo z_2$ and $\bo z_1\in R^k$
even if the true dimension $p\ne k$. Bootstrap sampling from a null distribution obeying the weaker assumptions such as $y\independent \bo z_2| \bo z_1$ and $\mathbb{E}(\bo z_2|\bo z_1)=\bo 0$ and $Cov(\bo z_2|y)=\bo I_{p-k}$  seems much more difficult to carry out and not developed here.  Sampling under our strong assumption is described in the following. \\

{\bf  Bootstrap strategy SIR:}\ \ {\it  Generate from the SIR model.
\begin{enumerate}
\item  Starting from $\bo X$,  find $\bar{\bo x}$ and $\widehat{\bo W}=(\widehat{\bo W}_1', \widehat{\bo W}_2')'$ where $\widehat{\bo W}_1\in \mathbb{R}^{k\times p}$ and \\  write   $\widehat{\bo Z}_i=(\bo X-\bo 1_n\bar{\bo x}')\widehat{\bo W}_i'$, $i=1,2$.
\item Let $(y^*,{\bo Z}_1^*)$ be a bootstrap sample of size $n$ from  $(y,\widehat{\bo Z}_1) $.
\item Let $\bo Z_2^*$ be a bootstrap sample of size $n$ from $\widehat{\bo Z}_2$.\\ (Bootstrap samples in 2 and 3  are independent)
\item Write  ${\bo Z}^*=(\bo Z_1^*, \bo Z_2^*)$.
\item Write $(\bo y^*, \bo X^*)=\left(\bo y^*,\widehat{\bo Z}^* (\widehat{\bo W}')^{-1} + \bo 1_n\bar{\bo x}'\right)$.
\end{enumerate}}

In other terms, the bootstrap null distribution $F_{k,n}$ at $(y,\bo x')'$ is now obtained as the average
\[
\frac 1{n^2}\sum_{i=1}^n\sum_{j=1}^n
\mathbb{I}\left(
\left(
  \begin{array}{c}
    \bo y'\bo e_i \\
    \widehat{\bo W}^{-1}
\left(
  \begin{array}{c}
    \widehat{\bo W}_1(\bo X-\bo 1_n\bar{\bo x}')' \bo e_i \\
    \widehat{\bo W}_2 (\bo X-\bo 1_n\bar{\bo x}')' \bo e_j \\
  \end{array}
\right)
     \\
  \end{array}
\right)
+ \left(
      \begin{array}{c}
        0 \\
        \bar{\bo x} \\
      \end{array}
    \right)
  \le \left(
      \begin{array}{c}
        y \\
        {\bo x} \\
      \end{array}
    \right)  \right)
\]
where the  $\bo e$'s are in $\mathbb{R}^{n}$.
As for PCA and ICA bootstrap strategies, let $\bo X_N^*$ be a sample of size $N$ from $F_{k,n}$ for which the null hypothesis $H_{0k}$ and our model
assumptions naturally hold true. Then
$NT_k(\bo X_N^*)\to_d \chi^2_{(p-k)(H-k-1)}$ and therefore, for large $n$, also the distribution of $nT_k(\bo X_n^*)$ can be approximated by the same distribution.
The estimated bootstrap $p$-value is obtained as in the previous cases.\\

\subsection{A simulation study}

To compare the bootstrap and asymptotic testing strategies, we consider the models  
\begin{description}
  \item[M1:] $y = z_1 (z_1 + z_2 +1) + \epsilon$ or
  \item[M2:] $y = z_1 / (0.5 + (z_2+1.5)^2) + \epsilon$
\end{description}
where  $\epsilon\sim N(0,0.25)$  and $\bo z\sim N_p(\bo 0, \bo I_p)$ are independent.  We then observe $y$ and $\bo x=\bo A\bo z+\bo b$ and the results are again the same for any choices of $\bo A$ and $\bo b$.
We use again the dimensions $p=6$ and $p=15$ and, in both models,  the dimension of the signal subspace $q=2$.\\

We compare our tests to the asymptotic test implemented in the dr package \citep{Weisberg2002} in R with the same limiting distribution but
an automated computation of the slices (the default number of slices is $\max\left\{8, p+3\right\}$). This test serves here  as a standard test choice  and is  called Asy1.
The asymptotic test Asy2 uses  the sample deciles of $y$ to separate the $H=10$ slices. In the bootstrap testing we use the sample deciles in the same way and choose $M=200$. Again, the choice $M=500$ would have only a minor effect on the accuracies of the  final rejection rate estimates.
 Table~\ref{Tab_SIR_M1} and Table~\ref{Tab_SIR_M2} report the observed rejection rates for $H_{01}$, $H_{02}$ (true) and $H_{03}$  at the level $\alpha=0.05$  over $N=2000$ repetitions
 for the models M1  and M2, respectively.

\begin{table}[ht]
\small
\centering
\begin{tabular}{@{\extracolsep{4pt}} rrrrrrrrrrr@{}}
  \hline
 \multicolumn{2}{c}{} & \multicolumn{3}{c}{$H_{01}$} & \multicolumn{3}{c}{$H_{02}$ (true)} & \multicolumn{3}{c}{$H_{03}$}\\ \cline{3-5} \cline{6-8} \cline{9-11}
p & n & Asy1 & Asy2 & Boot & Asy1 & Asy2 & Boot & Asy1 & Asy2 & Boot \\
  \hline
   6 &  100 & 0.171 &0.162 & 0.195  & 0.010 & 0.008 & 0.019& 0.001 & 0.000 & 0.004  \\
    &  200 & 0.531 &0.519 & 0.536  & 0.023 & 0.024 & 0.038& 0.001 & 0.001 & 0.008  \\
    &  500 & 0.987 &0.984 & 0.984  & 0.036 & 0.046 & 0.055& 0.001 & 0.001 & 0.009  \\
    & 1000 & 1.000 &1.000 & 1.000  & 0.055 & 0.050 & 0.057& 0.001 & 0.002 & 0.006  \\
    & 2000 & 1.000 &1.000 & 1.000  & 0.045 & 0.045 & 0.050& 0.002 & 0.000 & 0.004  \\
    & 5000 & 1.000 &1.000 & 1.000  & 0.055 & 0.053 & 0.051& 0.001 & 0.000 & 0.005  \\ \hline
  15 &  100 & 0.037 &0.060 & 0.086  & 0.002 & 0.002 & 0.010& 0.000 & 0.000 & 0.001  \\
   &  200 & 0.165 &0.255 & 0.283  & 0.009 & 0.015 & 0.026& 0.001 & 0.001 & 0.003  \\
   &  500 & 0.745 &0.855 & 0.861  & 0.028 & 0.028 & 0.039& 0.001 & 0.001 & 0.004  \\
   & 1000 & 0.995 &1.000 & 1.000  & 0.040 & 0.056 & 0.062& 0.001 & 0.001 & 0.005  \\
   & 2000 & 1.000 &1.000 & 1.000  & 0.044 & 0.045 & 0.055& 0.001 & 0.000 & 0.003  \\
   & 5000 & 1.000 &1.000 & 1.000  & 0.046 & 0.048 & 0.049& 0.001 & 0.002 & 0.004  \\
   \hline
\end{tabular}
\caption
{Rejection rates for $H_{01}$, $H_{02}$ (true)  and $H_{03}$ under the model M1 with two dimensions $p=6,15$ and for two asymptotic tests Asy1 and Asy2 and the bootstrap test Boot  with sample sizes $n=100,...,5000$. The target level for $H_{02}$ is 0.05.}
\label{Tab_SIR_M1}
\end{table}

\begin{table}[ht]
\small
\centering
\begin{tabular}{@{\extracolsep{4pt}} rrrrrrrrrrr@{}}
  \hline
 \multicolumn{2}{c}{} & \multicolumn{3}{c}{$H_{01}$} & \multicolumn{3}{c}{$H_{02}$ (true)} & \multicolumn{3}{c}{$H_{03}$}\\ \cline{3-5} \cline{6-8} \cline{9-11}
p & n & Asy1 & Asy2 & Boot & Asy1 & Asy2 & Boot & Asy1 & Asy2 & Boot \\
  \hline
   6 &  100 & 0.350 & 0.341 & 0.376  & 0.020 & 0.019 & 0.033 & 0.002 & 0.000 & 0.005  \\
    &  200 & 0.801 & 0.817 & 0.814  & 0.033 & 0.037 & 0.051 & 0.001 & 0.001 & 0.006  \\
    &  500 & 1.000 & 1.000 & 1.000  & 0.045 & 0.056 & 0.064 & 0.002 & 0.002 & 0.005  \\
    & 1000 & 1.000 & 1.000 & 1.000  & 0.034 & 0.045 & 0.048 & 0.001 & 0.001 & 0.007  \\
    & 2000 & 1.000 & 1.000 & 1.000  & 0.047 & 0.051 & 0.047 & 0.000 & 0.000 & 0.003  \\
    & 5000 & 1.000 & 1.000 & 1.000  & 0.062 & 0.045 & 0.047 & 0.002 & 0.000 & 0.002  \\ \hline
  15 &  100 & 0.093 & 0.149 & 0.183  & 0.007 & 0.009 & 0.021 & 0.000 & 0.000 & 0.002  \\
   &  200 & 0.350 & 0.487 & 0.518  & 0.017 & 0.027 & 0.037 & 0.001 & 0.001 & 0.005  \\
   &  500 & 0.966 & 0.991 & 0.990  & 0.035 & 0.038 & 0.047 & 0.001 & 0.001 & 0.004  \\
   & 1000 & 1.000 & 1.000 & 1.000  & 0.040 & 0.040 & 0.045 & 0.000 & 0.001 & 0.002  \\
   & 2000 & 1.000 & 1.000 & 1.000  & 0.037 & 0.047 & 0.052 & 0.000 & 0.002 & 0.004  \\
   & 5000 & 1.000 & 1.000 & 1.000  & 0.048 & 0.054 & 0.057 & 0.001 & 0.002 & 0.005  \\
   \hline
\end{tabular}
\caption
{Rejection rates for $H_{01}$, $H_{02}$ (true)  and $H_{03}$ under the model M2 with two dimensions $p=6,15$ and for two asymptotic tests Asy1 and Asy2 and the bootstrap test Boot  with sample sizes $n=100,...,5000$. The target level for $H_{02}$ is 0.05.  }
\label{Tab_SIR_M2}
\end{table}

The results are as expected and indicate for example that the rejection rates for true $H_{02}$ tend to be too small  for  small sample sizes. The bootstrap test reaches earlier the nominal level. The two asymptotic tests, Asy1 and Asy2  have different numbers of slices as well as different choices of slices;  Asy2  seems preferable in the considered cases.

\subsection{An example}

\begin{figure}[thp]
\centering
\includegraphics[width=0.9\textwidth]{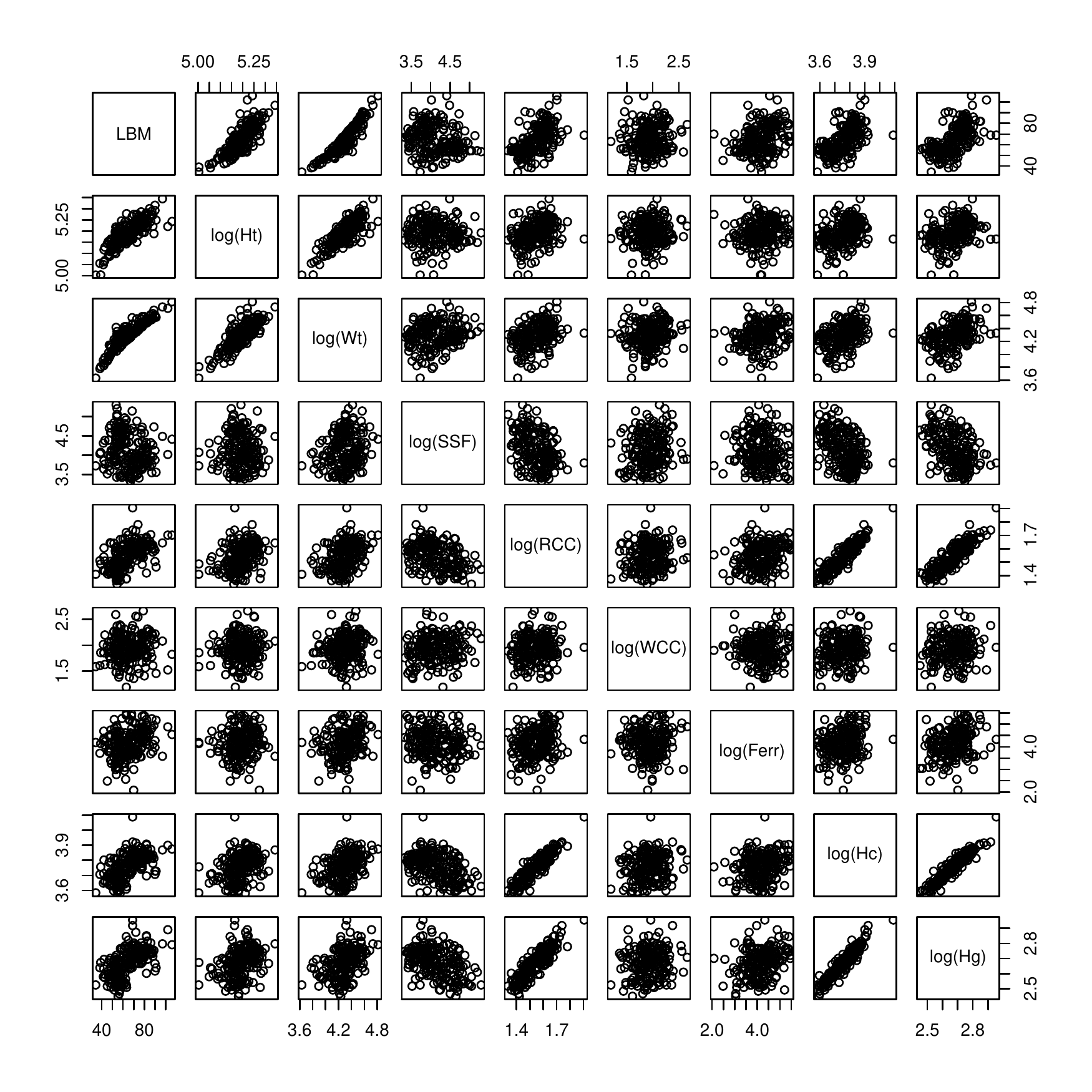}
\caption{Scatter plot matrix of the Australian athletes data.}
\label{Fig_AIS}
\end{figure}

We revisit the Australian Athletes data  available in the R package dr \citep{Weisberg2002}. The response variable $y$ is the lean body mass
the predictors in $\bo x$ are given by the logarithms of height (Ht),
weight (Wt), red cell count (RCC), white cell count (WCC), Hematocrit (Hc), Hemoglobin (Hg), plasma ferritin concentration (Ferr) and sum of skin folds (SSF). The same data was analysed e.g. by   \citet{Cook04} who developed tests of the hypothesis of no effect for a selected subset of  predictors. The data for all 202 athletes is shown in Figure~\ref{Fig_AIS} and the SIR eigenvalues are, rounding to two decimal places, $0.95, 0.21, 0.11, 0.07, 0.04, 0.02, 0.01$ and $0.00$.

\begin{table}[ht]
\centering
\begin{tabular}{rcccc}
  \hline
& $H_{00}$ & $H_{01}$  & $H_{02}$  &  $H_{03}$  \\
  \hline
SIR-I &    0.002 & 0.002 & 0.090 & 0.349 \\
Asymp &    0.000 & 0.001 & 0.121 & 0.458 \\
   \hline
\end{tabular}

\caption{ Australian Athletes Data: $p$-values for $H_{00}$-$H_{03}$ with two testing strategies.}
\label{Tab_AIS_pvals}
\end{table}

The observed $p$ values for successive testing of hypotheses $H_{00}$ to $H_{04}$  are reported in Table~\ref{Tab_AIS_pvals}. The number of bootstrap samples was $M=500$
and the bootstrap test as well as the asymptotic test suggest that the signal space has dimension two.
Note that the $p$-values of the asymptotic tests differ slightly from those in \citet{Cook04}, perhaps due to different number of slices and  different numbers of observations in slices.

\begin{figure}[thp]
\centering
\includegraphics[width=0.9\textwidth]{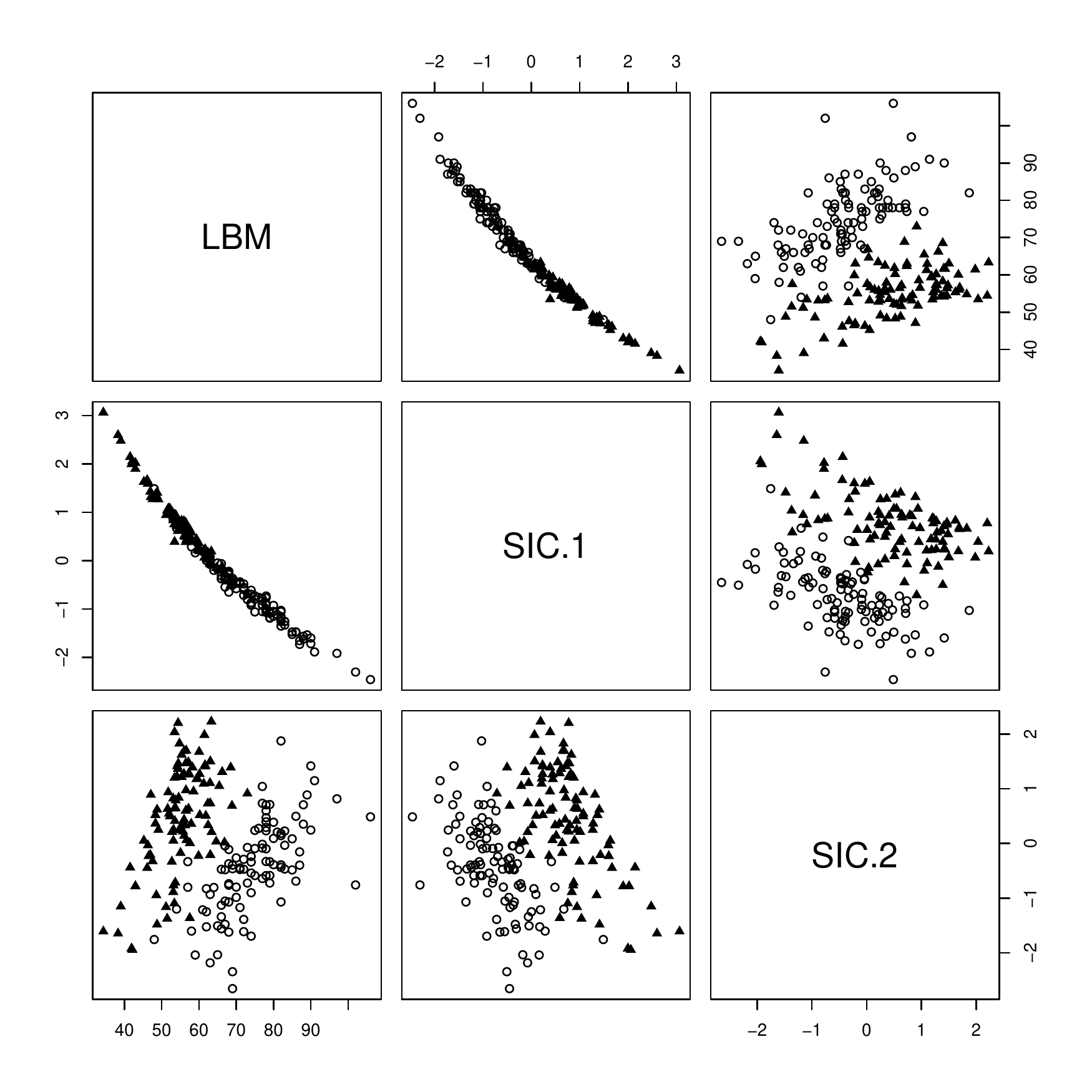}
\caption{Scatter plot matrix of the of the two selected SIR components against the response. Different plotting symbols have been used for men and women.}
\label{Fig_AIS_sir}
\end{figure}

The two signal components are plotted against the response in Figure~\ref{Fig_AIS_sir} where the plotting symbols differ for female and male athletes. The figure nicely shows that both components contain information about the response.

\section{Final remarks}\label{sec:final}
In this paper, we considered three  dimension reduction methods based on the use of a pair of sample matrices,   principal component analysis, fourth order blind identification  and sliced inverse regression,   and showed how first two moments of the eigenvalues of one matrix with respect to another can be used to test for signal (and noise) dimension. The asymptotic null distributions of the test statistics were given and novel bootstrap strategies were suggested for the testing problems. The asymptotic and bootstrap tests were compared in simulations studies and  in real data examples. These three methods serve here as examples and it is obvious that our approach can be extended to other pairs of scatter matrices tailored for the multivariate  semiparametric goodness-of-fit problems at hand, see e.g. \citet{Nordhausen:2011}.
The R code for all computations in the paper is available upon request from Klaus Nordhausen and almost all methods are implemented in the R package ICtest \citep{Nordhausen:2017b}.\\

 The estimation of the dimension and consistent estimates based on the sequential tests with different testing order strategies (bottom-up, top-down, divide
and conquer) are introduced in the paper as well. Wide simulation as well as theoretical studies in various contexts are necessary in the future to  compare
 the estimates here to other consistent estimates suggested in the literature \citep{YeWeiss03, Zhuetal10, LuoLi2016}.

\section{Appendix}\label{sec:appendix}

\subsection{Proofs for Section~\ref{sec:PCA} }

{\bf Proof of Lemma~\ref{L1:PCA}.}
Let $\widehat{\bo d} = (\hat{d}_{q+1}, \ldots, \hat{d}_p)$ denote the $r = p-q$ smallest ordered eigenvalues of $\widehat{\bo S}$ and
let $\widehat{\bo \delta} = (\hat{\delta}_{1}, \ldots, \hat{\delta}_{r})$ denote the ordered eigenvalues of $\widehat{\bo S}_{22}$.
Lemma 3.1  in Eaton and Tyler~(1991) then states that $\widehat{\bo d}-\widehat{\bo \delta}
= O_P\left(n^{-1}\right)$
and, applying Theorem 3.2 in Eaton and Tyler~(1991), $\widehat{\bo \delta} - d \bo 1_r = O_P\left(n^{-1/2}\right)$ then implies that $\widehat{\bo d} - d \bo 1_r = O_P\left(n^{-1/2}\right)$.
Setting $\bo P_r = \bo I_r -r^{-1} \bo 1_r \bo 1_r^\prime$, we then have $r\cdot T_q = \widehat{\bo d}^\prime \bo P_r \widehat{\bo d}
= (\widehat{\bo d} - d \bo 1_r)^\prime \bo P_r (\widehat{\bo d} - d \bo 1_r) $ and
$r\cdot s^2(\widehat{\bo S}_{22}) = \widehat{\bo \delta}^\prime \bo P_r \widehat{\bo \delta} =
(\widehat{\bo \delta} - d \bo 1_r)^\prime \bo P_r (\widehat{\bo \delta} - d \bo 1_r) $. Hence,
\[
r \left(T_q - s^2(\widehat{\bo S}_{22})\right) = 2(\widehat{\bo \delta} - d \bo 1_r)^\prime \bo P_r (\widehat{\bo d} - \widehat{\bo \delta})
+ (\widehat{\bo d} -\widehat{\bo \delta})^\prime \bo P_r (\widehat{\bo d} -\widehat{\bo \delta}),
\]
which is $O_P\left(n^{-3/2}\right) +O_P\left(n^{-2}\right) = O_P\left(n^{-3/2}\right)$. \\

{\bf Proof of Theorem~\ref{Th1:PCA}.}
By Lemma ~\ref{L1:PCA} it is sufficient to consider the limiting distribution of $n \cdot s^2(\widehat{\bo S}_{22})$. Let again $r=p-q$ and
$\bo Z_{22} = \sqrt{n}(\widehat{\bo S}_{22} - d \bo I_{r})/d$. Then
\[
nr \cdot s^2(\widehat{\bo S}_{22})/d^2 = n \cdot vec(\widehat{\bo S}_{22})^\prime {\bo \Gamma} vec(\widehat{\bo S}_{22})/d^2
= vec(\bo Z_{22})^\prime \bo \Gamma vec(\bo Z_{22}),
\]
where $\bo \Gamma = \bo I_{r^2} - r^{-1} vec(\bo I_{r})vec(\bo I_{r})^\prime$ is idempotent. The second identity follows since $\bo \Gamma vec(\bo I_r) = \bo 0$.
Under $H_{0q}$, $\bo Z_{22}  \to_d \mathrm{Z}$ with $vec(\mathrm{Z}) \sim N_{r^2}(\bo 0, \bo \Sigma)$,
where $\bo \Sigma = \sigma_1(\bo I_{r^2}+\bo K_{r,r}) + \sigma_2 vec(\bo I_{r})vec(\bo I_{r})^\prime$.
This implies
\[
nr\cdot s^2(\widehat{\bo S}_{22})/d^2 \to_d 2\sigma_1 \mathrm{z} ^\prime \mathrm{z}, \  \mbox{with} \
\mathrm{z} = \bo \Gamma vec(\mathrm{Z})/\sqrt{2\sigma_1} \sim N_{r^2}(\bo 0, \bo \Sigma_o),
\]
where
$$\bo \Sigma_0 =\bo\Gamma\frac{1}{2}(\bo I_{r^2}+\bo K_{r,r})\bo \Gamma 
=\frac 12 \left( \bo I_{r^2}+\bo K_{r,r}-\frac 2r vec(\bo I_{r})vec(\bo I_{r})^\prime \right) .$$
Now $\bo \Sigma_0$ is symmetric and idempotent with
rank$(\bo \Sigma_0) = (r^2+r-2)/2 = (r+2)(r-1)/2$, and so $\mathrm{z}^\prime\mathrm{z} \sim \chi^2_{(r+2)(r-1)/2}$ and the first part of the theorem follows.
The second part follows as $V_q$ is the minimum  of the variance over all $(p-q)$-subsets of the ordered eigenvalues of $\widehat{\bo S}$. The variance of the $p-q$ smallest eigenvalues, that is,  $T_q$ converges in probability to $0$, and the variance for any other ${p\choose q}-1$  choices of subsets converges in probability to a positive constant.

{\bf Proof of Theorem~\ref{Th2:PCA}.} (i) $T_k$ converges in probability to the variance of $p-k$ smallest eigenvalues which is positive for $k<q$. (ii) is given in the previous theorem. (iii) follows as,  for $k=q,...,p-1$,
\begin{eqnarray*}
T_k &=& \frac 1{2(p-k)^2}\sum_{i=k+1}^p \sum_{j=k+1}^p (\hat d_i-\hat d_j)^2\\
  &\le& \left(\frac {p-q}{p-k}\right)^2  \frac 1{2(p-q)^2}\sum_{i=q+1}^p \sum_{j=q+1}^p (\hat d_i-\hat d_j)^2 = \left(\frac {p-q}{p-k}\right)^2 T_q
\end{eqnarray*}

\subsection{Proofs for Section~\ref{sec:ICA} }

{\bf Proof of Lemma~\ref{L1:ICA}.}
This proof is similar to the proof of Lemma~\ref{L1:PCA}. Again set $r = p-q$. Rather than using the ordering of the roots given in Section \ref{sec:ICA},
let $\lambda_1, \ldots, \lambda_p$ denote the ordered eigenvalues of ${\bo R}$, and so for some $0 \le m \le q$, $\lambda_m > p+2$,
$\lambda_{m+1} = \cdots = \lambda_{m + r} = p+2$ and $\lambda_{m+r+1} < p+2$. Also, let
$\widehat{\bo \lambda} = (\hat{\lambda}_{m+1}, \ldots, \hat{\lambda}_{m+r})$ denote the $(m+1)th$ to $(m+r)th$ ordered eigenvalues
of $\widehat{\bo R}$ and let $\widehat{\bo \delta} = (\hat{\delta}_{1},  \ldots, \hat{\delta}_{r})$ denote the ordered eigenvalues of
$\widehat{\bo R}_{22}$. Again using Eaton and Tyler~(1991), applying its Lemma 3.1 twice gives $\widehat{\bo \lambda} -\widehat{\bo \delta} = O_P\left(n^{-1}\right)$
and applying its Theorem 3.2 gives $\widehat{\bo \lambda} - (p+2) \bo 1_p = O_P\left(n^{-1/2}\right)$.
Now, $r\cdot T_q = (\widehat{\bo \lambda} - (p+2) \bo 1_r)^\prime(\widehat{\bo \lambda} - (p+2) \bo 1_r) $ and
$r\cdot s^2(\widehat{\bo S}_{22}) = (\widehat{\bo \delta} - (p+2) \bo 1_r)^\prime (\widehat{\bo \delta} - (p+2) \bo 1_r) $. Hence,
\[
r \left(T_q - m_2(\widehat{\bo R}_{22})\right) = 2(\widehat{\bo \delta} - (p+2) \bo 1_r)^\prime (\widehat{\bo \lambda} - \widehat{\bo \delta})
+ (\widehat{\bo \lambda} -\widehat{\bo \delta})^\prime  (\widehat{\bo \lambda} -\widehat{\bo \delta}),
\]
which is $O_P\left(n^{-3/2}\right) +O_P\left(n^{-2}\right) = O_P\left(n^{-3/2}\right)$. \\

{\bf Proof of Theorem~\ref{Th1:ICA}}
By Lemma~\ref{L1:ICA} it is sufficient to consider the joint limiting distribution of $n(s^2(\widehat{\bo R}_{22}), m_1^2(\widehat{\bo R}_{22}))$.
Set again $r=p-q$.
The arguments for obtaining the limiting distribution of $n \cdot s^2(\widehat{\bo R}_{22})$ are analogous to those used in the proof of
Theorem~\ref{Th1:PCA}, and we use the same notation  but now with $\bo {Z}_{22} = \sqrt{n}(\widehat{\bo R}_{22} - (p+2) \bo I_{r})/(p+2)\to \mathrm{Z}$
with the property that $\bo U' \mathrm{Z} \bo U\sim \mathrm{Z}$ for all $\bo U\in \mathcal{O}^{r\times r}$. Then again  $vec(\mathrm{Z}) \sim N_{r^2}(\bo 0, \bo \Sigma)$,
where $\bo \Sigma = \sigma_1(\bo I_{r^2}+\bo K_{r,r}) + \sigma_2 vec(\bo I_{r})vec(\bo I_{r})^\prime$ with two population constants $\sigma_1$ and $\sigma_2$.
Using arguments analogous to those in the proof of Theorem~\ref{Th1:PCA}, we again obtain under the null hypothesis that
$nr\cdot s^2(\widehat{\bo R}_{22})/(p+2)^2 \to \chi^2_{(r+2)(r-1)/2}$.
Next, $r\sqrt{n}\cdot m_1(\widehat{\bo R}_{22}) = vec(\bo I_r)^\prime vec(\bo Z_{22}) \to_d vec(\bo I_r)^\prime vec( \mathrm{Z}) \sim N(0, \sigma^2)$,
with $\sigma^2 = vec(\bo I_r)^\prime \bo \Sigma vec(\bo I_r) = 2r\sigma_1 + r^2\sigma_2$. Thus $r^2n\cdot m_1^2 (\widehat{\bo R}_{22}) \to_d
\sigma^2\chi^2_1$. Finally, recall that, as in the proof of Theorem~\ref{Th1:PCA},  $n \cdot s^2(\widehat{\bo R}_{22})=vec(\bo Z_{22})^\prime \bo \Gamma vec(\bo Z_{22}) $
where  $\bo \Gamma vec(\bo I_r) = \bo 0$.
This establishes the independence of the limiting distributions of the component variables in  $(n\cdot s^2(\widehat{\bo R}_{22}), n\cdot m_1^2(\widehat{\bo R}_{22}))$, and consequently Theorem~\ref{Th1:ICA} follows with some constants $\sigma_1$ and $\sigma_2$. The values of $\sigma_1$ and $\sigma_2$ are derived in the Appendix in \citet{Nordhausen:2017}.

{\bf Proof of Theorem~\ref{Th2:ICA}.} (i) $T_k$ converges in probability to the sum of $p-k$ smallest eigenvalues of  $(\bo D-(p+2)\bo I_p)^2$ which is positive for $k<q$. (ii) is given in the previous theorem. (iii) follows as
\begin{eqnarray*}
T_{k} &=& \min_{\bo U\in \mathcal{O}^{p\times (p-k)}}  m_1\left(\bo U' (\widehat{\bo R}-(p+2)\bo I_p)^2 \bo U\right)\\
  &\le&  m_1\left((\bo 0, \bo I_{p-k}) (\widehat{\bo R}-(p+2)\bo I_p)^2(\bo 0, \bo I_{p-k})'  \right).
\end{eqnarray*}
and the result follows as,  for $k=q,...,p-1$, $(\bo 0, \bo I_{p-k})\widehat{\bo R}(\bo 0, \bo I_{p-k})'$ is a $(p-k)\times (p-k)$-submatrix of  $\widehat{\bo R}_{22}$
with the known limiting distribution.

\subsection{Proofs for Section~\ref{sec:SIR} }

{\bf Proof of Lemma~\ref{L1:SIR}.}
For $H \ge p$, let $\widehat{\bo \gamma} = (\hat{\gamma}_{q+1}, \ldots, \hat{\gamma}_p)'$ denote the $p-q$ smallest ordered singular values of $\widehat{\bo B}\bo Q'$.
When $q + 1 < H < p$,  we use the same notation while noting $\hat{\gamma}_{H+1} = \cdots = \hat{\gamma}_p = 0$. Likewise,
let $\widehat{\bo \eta} = (\hat{\eta}_{1}, \ldots, \hat{\eta}_{p-q})'$ denote the ordered singular values of $\widehat{\bo B}_{2} \bo Q_2'$.
Since $\sqrt{n}(\widehat{\bo B} - \bo B)\bo Q' = O_P(1)$, it follows respectively from Theorems 4.1 and 4.2 in \citet{EatonTyler94}
that $\widehat{\bo \gamma} - \widehat{\bo \eta} = O_P\left(n^{-3/4}\right)$ and $\widehat{\bo \gamma} = O_P\left(n^{-1/2}\right)$.
Next, observe that $(p-q) T_q = \widehat{\bo \gamma}^\prime\widehat{\bo \gamma}$ and
$(p-q) m_1(\widehat{\bo B}_{2}\bo Q_2'\bo Q_2 \widehat{\bo B}_{2}^\prime) = \widehat{\bo \eta}^\prime\widehat{\bo \eta}$.
Hence,
\[ (p-q)\{T_q - m_1(\widehat{\bo B}_{2}\bo Q_2'\bo Q_2 \widehat{\bo B}_{2}^\prime  )\} = 2\widehat{\bo \eta}^\prime(\widehat{\bo \gamma} - \widehat{\bo \eta}) +
(\widehat{\bo \gamma} - \widehat{\bo \eta})^\prime(\widehat{\bo \gamma} - \widehat{\bo \eta}), \]
which is $O_P\left(n^{-5/4}\right) +O_P\left(n^{-3/2}\right) = O_P\left(n^{-5/4}\right)$. \\

{\bf Proof of Theorem~\ref{Th1:SIR}.}
By Lemma~\ref{L1:SIR}, the limiting distributions of $n \cdot T_q$ and $n \cdot m_1(\widehat{\bo B}_{2}\bo Q_2'\bo Q_2 \widehat{\bo B}_{2}^\prime)$ are the same.
Let $\bo x_h^* \in \mathrm{R}^{p-q}$ refer to the last $p-q$ components of $ \mathbb{I}(y_i\in \mathbb{S}_h) \bo x \in \mathrm{R}^p$, $h=1,...,H$.
  Hence, under $H_{0q}$, $\bo x^* = \bo z_2$ is
independent of the response $y$.
Since $f_h \cdot \overline{\bo x}^*_{h} =  \frac 1{n}\sum_{i=1}^n \bo x^*_{(h),i}$, where $\bo x^*_{(h),i} =  \bo x^*_i \mathbb{I}(y_i\in \mathbb{S}_h)$,
with
$\mathbb{E}(\bo x^*_{(h)})= p_h \mathbb{E}(\bo x^*) = \bo 0$,
$Cov(\bo x^*_{(h)}) = p_h Cov(\bo x^*) = p_h \bo I_{p-q}$,
$Cov(\bo x^*_{(h)},\bo x^*_{(m)}) = \bo 0$ for $h \ne m$, and  $f_h \to_P p_h$,
it follows from the central limit theorem and from Slutsky's theorem that
$\sqrt{n}\left(\sqrt{f_1} ~ \overline{\bo x}^*_1, ~  \cdots ~, \sqrt{f_H} ~ \overline{\bo x}^*_H \right) \to_d \mathrm{Z}$,
where the elements of the $(p-q) \times H$ random matrix $\mathrm{Z}$ are i.i.d.\ $N(0,1)$.

 Since $\widehat{\bo S}_1 \to_p \bo I_p$ and $\overline{\bo x}^* = \sum_{h=1}^H f_h \overline{\bo x}^*_h$,
we obtain $\sqrt{n}\cdot \widehat{\bo B}_{2}\bo Q_2' \to_d \mathrm{Z} (\bo I_H - \bo \pi \bo \pi^\prime)\bo Q_2^\prime$ with $\bo \pi^\prime = (\sqrt{p_1}, \ldots, \sqrt{p_H})$. Hence
$n\cdot\widetilde{\bo B}_2\widetilde{\bo B}_2^\prime \to_d \mathrm{Z} \bo P \mathrm{Z}^\prime$, where
$\bo P = (\bo I_H - \bo \pi \bo \pi^\prime) \bo Q_2^\prime \bo Q_2 (\bo I_H - \bo \pi \bo \pi^\prime)$.
It is shown below that $\bo P$ is idempotent with rank $H-q-1$, which implies $\mathrm{Z} \bo P \mathrm{Z}^\prime \sim Wishart_{p-q}(H-q-1,\bo I_{p-q})$, and consequently,
$n \cdot tr(\widehat{\bo B}_{2}\bo Q_2'\bo Q_2 \widehat{\bo B}_{2}^\prime) \to_d tr(\mathrm{Z} \bo P \mathrm{Z}^\prime) \sim \chi^2_{(p-q)(H-q-1)}$.

To complete the proof,  note that since $\bo \mu = \bo 0$, it follows that
$\bo B \bo \pi = \bo 0$ and hence $\bo Q_1 \bo \pi = \bo 0$. Also, since $\bo I_H - \bo \pi \bo \pi^\prime$ is idempotent with rank $H-1$, we have
\[ \bo I_H - \bo \pi \bo \pi^\prime = (\bo I_H - \bo \pi \bo \pi^\prime)\bo Q' \bo Q (\bo I_H - \bo \pi \bo \pi^\prime)
=\bo Q_1'\bo Q_1 + \bo P, \]
which implies $\bo P$ is idempotent with rank $H-q-1$.

\section{Acknowledgements}
The authors wish to thank the associate editor and the referees for helpful and valuable comments and suggestions on an earlier version
of this paper. This article is to some extent based on Hannu Oja's presentation at {\it the 2nd Workshop on goodness-of-fit and change-point problems}
in Athens, September 2015. The research of Klaus Nordhausen and Hannu Oja  was partially supported by the Academy of
Finland (grant 268703). David E. Tyler's research was partially supported by the National Science Foundation
Grant No. DMS-1407751. Any opinions, findings and conclusions or recommendations expressed in this material
are those of the author(s) and do not necessarily reflect those of the National Science Foundation.


\begin{thebibliography}{99.}



\bibitem [Beran and Srivastava(1985)] {BeranSrivastava85}
Beran, R. and Srivastava, M. S.: Bootstrap tests and confidence regions for functions of
a covariance matrix. {\it Annals of Statistics}, {\bf 13}, 95--115 (1985).
Correction: {\it  Annals of Statistics}, {\bf 15}, 470--471 (1987).


\bibitem [Bilodeau and Brenner(1999)] {BB99}
Bilodeau, M. and Brenner, D. (1999). {\it Theory of multivariate statistics}. Springer, New York.



\bibitem [Blanchard et al.(2005)] {Blanchard05}
Blanchard, G., Sugiyama, M.,   Kawanabe, M., Spokoiny, V. and K.-R. M\"uller, K.-R.:
 Non-Gaussian component analysis: a semi-parametric framework
for linear dimension reduction. In {\it Advances in Neural Information Processing
Systems}, 131--138 (2005)


\bibitem [Bura  and Cook(2001)] {BuraCook01}
Bura, E. and Cook, R.D. (2001).
Extending sliced inverse regression:
the weighted chi-squared test. {\it Journal of the American Statistical Association}, {\bf
96}, 996--1003.


\bibitem[Bura and Yang(2011)]{BuraYang11}
Bura, E. and Yang, J. (2011)
Dimension estimation in
sufficient dimension reduction: a unifying approach.
{\em Journal of Multivariate Analysis}, {\bf 102}, 130--142.


\bibitem[Cardoso(1989)]{Cardoso89}
{Cardoso, J.~F.} (1993).
\newblock Source separation using higher order moments.
\newblock {\em Proceedings of IEEE International Conference on Acoustics, Speech and Signal Processing}, {\bf  4}, 2109--2112.

\bibitem[Caussinus and Ruiz-Gazen(1993)]{CaussinusRuizGazen93}
{ Caussinus, H. and Ruiz-Gazen, A.} (1993).
 Projection pursuit and
generalized principal component analysis.
{\em New Directions in
Statistical Data Analysis and Robustness}, eds. by S. Morgenthaler,
E. Ronchetti, and W.A. Stahel, 328--332, Basel: Birkh\"auser.

\bibitem [Cook(2004)] {Cook04}
Cook, R. D. (2004). Testing predictor contributions in sufficient dimension reduction.
{\it Annals of Statistics}, {\bf 3},  1062--1092


\bibitem[Cook and Weisberg(1991)]{CookWeisberg91}
{ Cook, R.~D. and Weisberg, S.} (1991).
Sliced inverse regression for dimension reduction: comment.
 {\em Journal of the American Statistical Association},
{\bf 86}, 328--332.

\bibitem [Croux and Haesbrock(2000)] {CH00}
Croux, C. and Haesbrock, G. (2000). Principal component analysis
based on robust estimators of the covariance or correlation matrix:
Influence functions and efficiencies. {\it Biometrika}, {\bf
87},603--618.

\bibitem [Daudin et al.(1988)] {Daudin88}
Daudin, J., Duby, C., and Trecourt, P. (1988). Stability of principal component analysis stud-
ied by the bootstrap method. {\it Statistics: A Journal of Theoretical and Applied Statistics},
{\bf 19}, 241--258.

\bibitem[Dray(2008)]{Dray08}
Dray, S.T. (2008).  On the number of principal components: A test of dimensionality
based on measurements of similarity between matrices. {\it Computational Statistics \& Data Analysis}, {\bf 52},  2228--2237.

\bibitem[D\"umbgen et~al.(2016)]{Duembgen:2016}
D\"umbgen, L., Nordhausen, K. und Schuhmacher, H. (2016)
New algorithms for M-estimation of multivariate scatter and location.
{\em Journal of Multivariate Analysis}, {\bf 144} 200--217.


\bibitem[Eaton and Tyler(1991)]{EatonTyler91}
Eaton, M.L. and Tyler, D.E. (1991). On Wielandt's inequality and its application to the asymptotic distribution of the
eigenvalues of a random symmetric matrix.
{\it Annals of  Statistics},  {\bf 19},  260"1¤7271.

\bibitem[Eaton and Tyler(1994)]{EatonTyler94}
Eaton, M. L. and Tyler, D.E. (1994). The asymptotic distribution of singular values with application
to canonical correlations and correspondence analysis. {\it Journal of Multivariate Analysis}, {\bf 50},
238--264 (1994)


\bibitem [Hall  and Wilson(1991)] {HallWilson91}
Hall, P. and Wilson, S.R. (1991).
Two guidelines for bootstrap hypothesis testing. {\it Biometrics}, {\bf  47}, 757--762


\bibitem[Hampel et al.(1986)]{hampeletal:1986}
Hampel, F.R., Ronchetti, E.M., Rousseeuw, P.J., Stahel, W.J. (1986).
{\it Robust Statistics: the Approach Based on Influence Functions}.
Wiley, New York.


\bibitem[Hettmans\-perger and Randles(2002)]{hettmanspergerandrandles:2002}
Hettmansperger, T.P. and  Randles, R.H. (2002).
A practical affine equivariant  multivariate median,
{\it Biometrika}, {\bf 89}, 851--860.



\bibitem [Huber(1981)] {H81}
Huber, P.J. (1981). {\it Robust Statistics}. Wiley, New York.






\bibitem[Ilmonen et al.(2012)]{ISO12}
Ilmonen, P., Serfling, R., and Oja, H. (2012).
Invariant coordinate selection (ICS) functionals.
{\it International Statistical Review}, {\bf 80}, 93--110.

\bibitem[Jackson(1993)]{Jackson93}
Jackson, D.A. (1993). Stopping rules in principal components analysis: A comparison of heuristical and statistical
approaches. {\it Ecology}, {\bf 74}, 2204--2214.

\bibitem[Jolliffe(2002)]{Jolliffe02}
Jolliffe, I. T. (2002). {\it Principal component analysis}. Springer, Berlin.

\bibitem[Kankainen et al.(2007)]{Kankainen07}
Kankainen, A., Taskinen, S. and Oja, H. (2007). Tests of multinormality based on location vectors and scatter matrices.
{\it Statistical Methods \& Applications}, {\bf 16}, 357--379.

\bibitem[Kelker(1970)]{kelker70}
Kelker, D. (1970).
Distribution theory of spherical distributions and a location-scale parameter generalization.
{\it Sankhya Ser. A}, {\bf 32},  419--430.

\bibitem[Lane(2016)]{Lane16}
Lane, D. M. (2016).
The assumption of sphericity in repeated-measures designs: What it means and what to do when it is violated.
{\it The Quantitative Methods for Psychology}, {\bf 12}, 114--122.



\bibitem[Li(1991)]{Li91}
Li, K. C. (1991): Sliced Inverse Regression for Dimension Reduction,
{\it  Journal of the American Statistical Association}, {\bf 86}, 316"1¤7342.


\bibitem[Li(1992)]{Li92}
{ Li, K.~C.} (1992).
On principal Hessian directions for data visualization and dimension reduction: another application of Stein's lemma.
{\em Journal of the American Statistical Association},
{\bf 87}, 1025--1039.

\bibitem[Liski et al.(2014)]{Liski2014}
Liski, E., Nordhausen, K. and Oja, H. (2014). Supervised Invariant Coordinate Selection.
{\it Statistics: A Journal of Theoretical and Applied Statistics}, {\bf 48}, 711--731.

\bibitem[Luo and Li(2016)]{LuoLi2016}
Luo, W. and Li, B.(2016).
Combining eigenvalues and variation of eigenvectors for
order determination.
{\it  Biometrika}, {\bf 103}, 875"1¤7-887.


\bibitem[Ma and Zhu(2013)]{MaZhu13}
Ma, Y. and Zhu, L. (2013). A Review on Dimension Reduction.
{\it  International Statistical  Reviews}, {\bf 81}.

\bibitem[Maronna(1976)]{maronna:1976}
Maronna, R.A. (1976). Robust M-estimators of multivariate location and scatter.
{\it Annals of Statistics}, {\bf 4}, 51--67.


\bibitem[Nordhausen and Tyler(2015)]{Nordhausen:2015}
Nordhausen, K. and Tyler, D.E. (2015): A Cautionary Note on Robust Covariance Plug-in Methods. \textit{Biometrika}, \textbf{102}, 573--588,

\bibitem[Nordhausen et~al.(2011)]{Nordhausen:2011}
Nordhausen, K., Oja, H. and Ollila, E. (2011): Multivariate Models and
 the First Four Moments. In Hunter, D.R., Richards, D.S.R. and
 Rosenberger, J.L. (editors) {\it ``Nonparametric Statistics and Mixture
 Models:  A Festschrift in Honor of Thomas P. Hettmansperger''}, 267--287,
 World Scientific, Singapore.

\bibitem[Nordhausen et~al.(2008)]{Nordhausen_b:2008}
{ Nordhausen, K., Oja, H. and Tyler, D.~E.} (2008).
Tools for exploring multivariate data: the package ICS.
{\em Journal of Statistical Software},
{\bf 28}, 1--31. 

\bibitem[Nordhausen et~al.(2017)]{Nordhausen:2017}
Nordhausen, K., Oja, H., Tyler, D.E. and Virta, J. (2017).
 Asymptotic and bootstrap tests for the dimension of the non-Gaussian subspace.{\em IEEE Signal Processing Letters}, {\bf 6}, 887--891.


\bibitem[Nordhausen et~al.(2017b)]{Nordhausen:2017b}
Nordhausen, K., Oja, H., Tyler, D.E. and Virta, J. (2017b).
 ICtest: Estimating and Testing the Number of Interesting Components
  in Linear Dimension Reduction. R package version 0.3.
  https://CRAN.R-project.org/package=ICtest.

\bibitem[Nordhausen et~al.(2015)]{ICSNP:2015}
Nordhausen, K., Sirki\"a, S., Oja, H. and Tyler, D.E. (2015).
ICSNP: Tools for Multivariate Nonparametrics. R package version 1.1-0.
https://CRAN.R-project.org/package=ICSNP.



\bibitem [Oja et~al.(2006)]{Oja:2006}
{ Oja, H., Sirki\"a, S. and Eriksson, J.} (2006).
 Scatter matrices and independent component analysis.
 {\em Austrian Journal of Statistics},  {\bf 35}, 175--189.

\bibitem [Saliban-Barrera et~al.(2005)]{saliban:2005}
Salibian-Barrera, M., Van Aelst, S., and Willems, G. (2005).
PCA based on multivariate MM-estimators with fast and robust bootstrap.
{\em Journal of the American Statistical Association} {\bf 101}, 1198--1211.

\bibitem [Saliban-Barrera and Zamar(2002)]{saliban:2002}
Saliban-Barrera, M. and Zamar, R.H. (2002).
Bootstrapping robust estimates of regression.
{\em The Annals of Statistics}, {\bf 30}, 556--582.




\bibitem[Schott(2006)]{schott06}
Schott, J. R. (2006).
A high-dimensional test for the equality of the
smallest eigenvalues of a covariance matrix.
{\it Journal of Multivariate Analysis}, {\bf 97}, 827--843.


\bibitem[Taskinen and Oja(2016)]{Taskinen:2016}
Taskinen, S. and Oja, H. (2016). Influence functions and efficiencies of k-step Hettmansperger-Randles estimators. In R. Liu and J.W. McKean, eds. Robust Rank-Based and Nonparametric Methods, Springer, Heidelberg, pp. 189--207.

\bibitem [Theis et~al.(2011)]{Theis:2011}
Theis, F. J., M. Kawanabe, and K.-R. M\"uller (2011). Uniqueness of non-
gaussianity-based dimension reduction. {\it IEEE Transactions on Signal
 Processing}, {\bf  59}, 4478--4482.

\bibitem[Tyler(1982)]{tyler:1982}
 Tyler, D.E. (1982).
 Radial estimates and the test for sphericity.
{\it Biometrika}, {\bf 69}, 429--436.

\bibitem[Tyler(1983)]{tyler:1983}
 Tyler, D.E. (1983).
The asymptotic distribution of principal component roots under local alternatives to
multiple roots.
{\it Annals of Statistics}, {\bf 11}, 1232--1242.


\bibitem[Tyler(1987)]{tyler:1987}
 Tyler, D.E. (1987).
A distribution-free M-estimator of multivariate scatter.
{\it Annals of Statistics}, {\bf 15}, 234--251.

\bibitem[Tyler et al.(2009)]{TCDO09}
Tyler, D., Critchley, F., D\"umbgen, L. and Oja, H. (2009).
Invariant coordinate selection.
{\it Journal of Royal Statistical Society B}, {\bf 71},  549--592.

\bibitem[Vieira(2012)]{Vieira12}
Vieira, V.M.N.C.S. (2012).  Permutation tests to estimate significances on Principal Components Analysis.
{\it  Computational Ecology and Software}, {\bf 2}, 103--123.

\bibitem[Weisberg(2002)]{Weisberg2002}
{Weisberg, S.} (2002).
\newblock Dimension reduction regression in R.
\newblock {\em Journal of Statistical Software}
7, 1--22. 


\bibitem[Ye and Weiss(2003)]{YeWeiss03}
Ye, Z. and  Weiss, R.E. (2003). Using the bootstrap to select one of a new class of dimension reduction methods.
{\it Journal of the American Statistical  Association}, {\bf 98}, 968--979.

\bibitem[Zhu et al.(2006)]{Zhuetal06}
Zhu, L., Miao, B. and Peng, H. (2006).
On sliced inverse regression with
high-dimensional covariates.
{\it  Journal of the American Statistical Association}, {\bf 101}, 630--643.

\bibitem[Zhu et al.(2010)]{Zhuetal10}
Zhu, L., Wang, T., Zhu, L and Ferr\'e, L. (2010).
Sufficient dimension reduction through
discretization-expectation estimation.
{\it Biometrika}, {\bf 97}, 295--304.

\end{thebibliography}
\end{document}